\documentclass[twocolumn]{aastex62}
\usepackage{amsmath}
\usepackage{amsfonts}

\submitjournal{ApJL}
\received{February 13, 2023}
\revised{April 3, 2023}
\accepted{April 5, 2023}

\usepackage[]{color}

\begin{document}

\title{Three-dimensional dynamics of strongly twisted magnetar magnetospheres: Kinking flux tubes and global eruptions}
\shorttitle{Kinking 3D flux tubes and global eruptions}
\shortauthors{Mahlmann et al.}

\correspondingauthor{J. F. Mahlmann}
\email{mahlmann@princeton.edu}

\author[0000-0002-5349-7116]{J. F. Mahlmann}
\affiliation{Department of Astrophysical Sciences, Peyton Hall, Princeton University, Princeton, NJ 08544, USA}

\author[0000-0001-7801-0362]{A. A. Philippov}
\affiliation{Department of Physics, University of Maryland, College Park, MD 20742, USA}

\author[0000-0001-5869-8542]{V. Mewes}
\affiliation{National Center for Computational Sciences, Oak Ridge National Laboratory, Oak Ridge, TN 37831, USA}

\author[0000-0002-7301-3908]{B. Ripperda}
\altaffiliation[NASA Hubble Fellowship Program, Einstein Fellow]{}
\affiliation{School of Natural Sciences, Institute for Advanced Study, Princeton, NJ 08540, USA}
\affiliation{Department of Astrophysical Sciences, Peyton Hall, Princeton University, Princeton, NJ 08544, USA}
\affiliation{Center for Computational Astrophysics, Flatiron Institute, New York, NY 10010, USA}

\author[0000-0002-0491-1210]{E. R. Most}
\affiliation{Princeton Center for Theoretical Science, Jadwin Hall, Princeton University, Princeton, NJ 08544, USA}
\affiliation{Princeton Gravity Initiative, Jadwin Hall, Princeton University, Princeton, NJ 08544, USA}
\affiliation{School of Natural Sciences, Institute for Advanced Study, Princeton, NJ 08540, USA}

\author[0000-0002-5951-0756]{L. Sironi}
\affiliation{Department of Astronomy and Columbia Astrophysics Laboratory, Columbia University, New York, NY 10027, USA}


\begin{abstract}

The origins of the various outbursts of hard X-rays from magnetars (highly magnetized neutron stars) are still unknown. We identify instabilities in relativistic magnetospheres that can explain a range of X-ray flare luminosities. Crustal surface motions can twist the magnetar magnetosphere by shifting the frozen-in footpoints of magnetic field lines in current-carrying flux bundles. Axisymmetric (2D) magnetospheres exhibit strong eruptive dynamics, as to say, catastrophic lateral instabilities triggered by a critical footpoint displacement of $\psi_{\rm crit}\gtrsim\pi$. In contrast, our new three-dimensional (3D) twist models with finite surface extension capture important non-axisymmetric dynamics of twisted force-free flux bundles in dipolar magnetospheres. Besides the well-established global eruption resulting (as in 2D) from lateral instabilities, such 3D structures can develop helical, kink-like dynamics, and dissipate energy locally (confined eruptions). Up to $25\%$ of the induced twist energy is dissipated and available to power X-ray flares in powerful global eruptions, with most of our models showing an energy release in the range of the most common X-ray outbursts, $\lesssim 10^{43}$erg. Such events occur when significant energy builds up deeply buried in the dipole magnetosphere. Less energetic outbursts likely precede powerful flares, due to intermittent instabilities and confined eruptions of a continuously twisting flux tube. Upon reaching a critical state, global eruptions produce the necessary Poynting-flux-dominated outflows required by models prescribing the fast radio burst production in the magnetar wind -- for example, via relativistic magnetic reconnection or shocks.

\end{abstract}

\section{Introduction}

Bursts and flares of hard X-rays are a common feature in magnetar observations \citep{Gogus1999,Gogus2000,Gogus2001,Rea2009,Rea2010,Kaspi_2017ARAaA..55..261,Esposito2020}. Due to their short duration as well as a rapid rise time and variability, the origins of these events must be outside of the neutron star, in the strongly magnetized, relativistic magnetosphere. Axisymmetric (2D) magnetospheric instabilities capable of triggering X-ray transients were simulated in the infinitely magnetized limit of force-free electrodynamics \citep[FFE,][]{Parfrey2013,Mahlmann2019,yuan2020,Sharma2023}. \citet{Carrasco:2019aas} evaluated the three-dimensional (3D) bursting dynamics of non-axisymmetric shears. These models show that twisting magnetic field lines by crustal surface motions can trigger a significant dynamical rearrangement of the magnetosphere, in 2D generally resulting in reconnection events and global eruptions. Up to now, the full 3D nature of such magnetospheric instabilities has not been taken into account, specifically the potential growth of non-axisymmetric plasma instabilities such as the helical (kink) mode. This Letter studies magnetospheric instabilities in 3D and identifies how helical, kink-like instabilities, the onset of magnetic reconnection, and the magnetic pressure imbalance of flux tubes in the dipole magnetosphere depend on non-axisymmetric distributions of magnetospheric shear induced by surface motions. 

The physical processes examined in this paper resemble, in their essence, those of solar magnetic field dynamics of coronal mass ejections \citep[CMEs, e.g.,][]{Chen2011,Manchester2017}. Though also addressed analytically \citep[see][]{Isenberg2007}, flux ropes and the field line configurations emerging from the injection of twist and helicity by surface motions are a common subject of simulations \citep[e.g.,][]{Amari2003,Gerrard2004,Torok2005,Torok2010,Gordovskyy2011,Gordovskyy2014,Pinto2016,Ripperda2017,Ripperda2017a,Sauppe2018}. This Letter shows that similarly rich dynamics are expected in the corona around magnetars, a region of active plasma processes in highly relativistic magnetar magnetospheres \citep[e.g.,][]{Beloborodov2013}. 

Recent observations associated some of the magnetar X-ray activity with the production of powerful radio bursts. The two notable examples are outbursts from the galactic magnetar \object{SGR J1935+2154}, and \object{1E 1547.0-5408} \citep{Bochenek2020,Chime2020,Israel2021,Li2021,Ridnaia2021,Younes2021}. These multiwavelength observations require theoretical models in the global context. While any release of the magnetic energy in the immediate magnetar vicinity produces X-ray emission, radio bursts are likely related to the outflow dynamics around or beyond the light cylinder \citep[see, e.g.,][]{Lyubarsky2021}. However, by far, not all the X-ray emission episodes are associated with radio events \citep[see, e.g.,][]{Kaspi_2017ARAaA..55..261,Esposito2020}. This Letter identifies magnetospheric instabilities with potential implications for consolidating some of the transient magnetar phenomenologies.

This Letter is organized as follows. Section~\ref{sec:methodology} describes our numerical method and initial conditions. We present our results in Section~\ref{sec:results}. Section~\ref{sec:axisymmetric} explores magnetospheres with an axisymmetric twist, 
Section~\ref{sec:openversuslocal} discusses the instabilities of 3D twisted magnetospheres. We scan the parameter space of magnetospheric twist to determine its impact on the magnetospheric dynamics (Section~\ref{sec:hotspotcharacteristics}). Section~\ref{sec:longterm} then shows the long-term evolution of three selected twisting geometries. We discuss how different magnetospheric eruption scenarios can result in different magnetar outbursts in Section~\ref{sec:discussion} and summarize our conclusions in Section~\ref{sec:conclusion}. 

\section{Methodology}
\label{sec:methodology}

\begin{figure}
 \centering
 \includegraphics[width=0.48\textwidth]{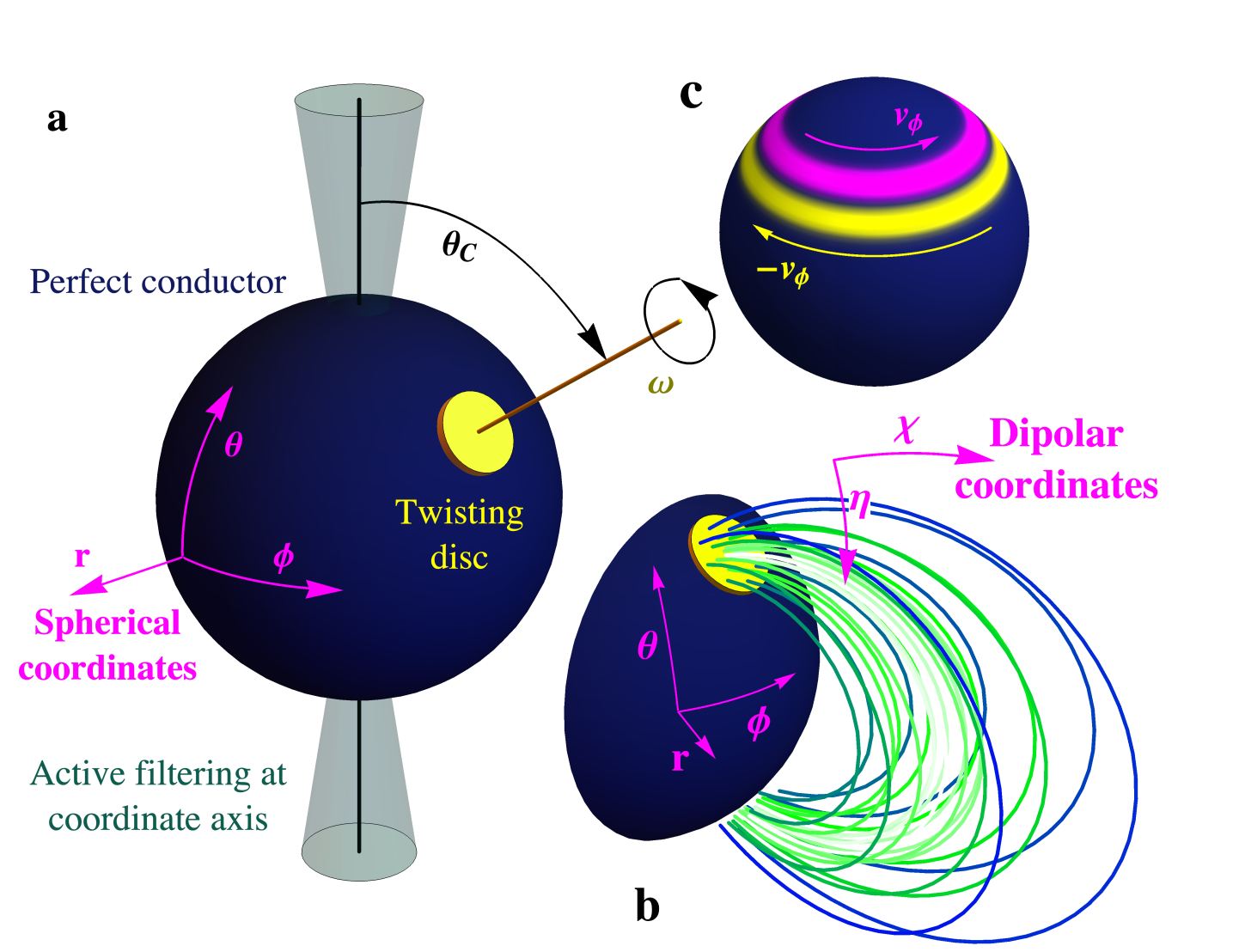}
 \caption{Schematic visualization of the simulation setup. The stellar surface is modeled as a perfect conductor (Section~\ref{sec:perfectconductorboundary}). In 3D, a rotating disk is centered at the colatitude $\theta_{\rm C}$, with angular extent $\theta_{\rm T}$ and angular velocity $\omega_0$ (a). The flux bundle (b) can be described in spherical $(r,\theta,\phi)$ or dipolar coordinates $(\eta,\chi,\phi)$. The 2D equivalent setup consists of two rings shearing the field lines in opposite directions (c).}
\label{fig:Figure1}
\end{figure}

The global dynamics of nearly force-free magnetar magnetospheres can be modeled approximately using relativistic FFE \citep[see, e.g.,][]{gruzinov1999,Blandford2002,Spitkovsky2006}. Studying the slow evolution of field topologies toward rapidly growing instabilities requires highly accurate numerical methods \citep[see][]{Parfrey2012,Carrasco2017,Most2020,yuan2020,ripperda2021}. For this work, we overcame two specific challenges. First, the motion of magnetic field line footpoints at the stellar surface has to be well resolved. Spherical meshes are favorable because they allow an accurate representation of the perfectly conducting magnetar surface. However, such meshes limit the possible integration time step to prohibitively small values because of small cell volumes at the coordinate axis. Second, the numerical diffusivity must be small for many dynamical time scales of the system, to avoid the contamination of physical instabilities with discretization noise. As a solution to these issues, we employ a high-order FFE method with optimized hyperbolic/parabolic cleaning parameters \citep{Mahlmann2020b,Mahlmann2020c,Mahlmann2021} that vastly benefits from the \textsc{Carpet} driver \citep{Goodale2002a,Schnetter2004}. Its extension to spherical coordinates \citep{Mewes2018,Mewes2020} is supported by the infrastructure of the \textsc{Einstein Toolkit} \citep{Loeffler2012,yosef_zlochower_2022_6588641}\footnote{\url{http://www.einsteintoolkit.org}}. 

The numerical simulations presented in this work target the highly magnetized magnetar magnetosphere, approximated by a dipole field configuration anchored to a perfectly conducting sphere (see Figure~\ref{fig:Figure1}, panels a/b). We model the force-free magnetar magnetosphere in the uniformly spaced, discrete domain (indicated by a bar) $\bar{r}\times\bar{\theta}\times\phi=\left[R_*,\bar{r}_{\rm max}\right]\times\left[0,\pi\right]\times\left[0,2\pi\right]$, where $R_*$ denotes the radius of the magnetar surface and $\bar{r}_{\rm max}$ is far away from the central object, covering at least twice the distance that signals with the speed of light travel during the simulation time. The employed resolutions are combinations of the grid spacing $\Delta\bar{r}=R_*/N_r$ with $N_r \in\left[32,64\right]$, $\Delta\bar{\theta}=\pi/N_\theta$ with $N_\theta\in\left[100,200\right]$, and $N_\phi=2N_\theta$. 2D reference cases employ the same mesh setup in $\bar{r}\times\bar{\theta}$ but discard any $\phi$ dependence. The radial and polar coordinates are rescaled to focus high resolutions close to the star and in the equatorial region. Specifically, we introduce the radial coordinate $r$ as equidistant with $\Delta r=\Delta\bar{r}$ for $\bar{r}\lesssim 40 R_*$ and increase $\Delta r$ by a factor of $a = 1.001$ in each grid point along the radial direction beyond this point. We increase the effective resolution in the equatorial plane to reduce the leakage of twist across magnetic field lines around the equator (see benchmarks in Appendix~\ref{sec:diffusivity}). Following \citet[][]{Porth2017}, we rescale the angular coordinate as
\begin{align}
 \theta\rightarrow \bar{\theta}+\frac{f_\theta}{2}\sin\left(2\bar{\theta}\right),
 \label{eq:angularscaling}
\end{align}
where $f_\theta$ determines the refinement factor at the equator (i.e., the coarsening factor at the coordinate axis), and our simulations use $f_\theta=2$. In our setups, only slow large-scale dynamics occur close to the coordinate axis; therefore, we can assume quasi-axisymmetry and filter high-frequency noise in all field quantities close to the coordinate singularity. By employing this technique, we can use a time step that is a factor of $30$ greater than that prescribed by the CFL condition for the spherical integration \citep[][see Appendix~\ref{sec:filtering} for further details]{Courant1928}.

For all simulations shown in this Letter, we initialize a dipolar magnetic field in spherical coordinates:
\begin{align}
\begin{split}
 \mathbf{B}_d&=\mu\left(\frac{2\cos\theta}{r^3},\frac{\sin\theta}{r^4},0\right),\\
 B_d&=\frac{\mu}{r^3}\sqrt{3\cos^2\theta+1}.
\end{split}
 \label{eq:dipoleinit}
\end{align}
Here, $\mu=B_* R^3_*/2$ is the magnetic moment of the star, and $B_*$ is the surface magnetic field strength. We twist selected field lines enclosed by a small disk on the surface \citep[Figure~\ref{fig:Figure1}, panel a, similar to][]{Carrasco:2019aas,Yuan2022}, prescribing a drift velocity $\mathbf{v}_{\rm T}$, centered at colatitude $\theta_{\rm C}$ with an angular extension $\theta_{\rm T}$. The corresponding electric field is derived from Ohm's law of ideal (MHD) plasma, $\mathbf{E}=-\mathbf{v}_{\rm T}\times\mathbf{B}/c\;$. We initialize the twist $v^{\tilde{\phi}}=\omega(\tilde{\theta})$ as an axisymmetric polar velocity field in the associated coordinates $(r,\tilde{\theta},\tilde{\phi})$:
\begin{align}
 \omega(\tilde{\theta})=\frac{\omega_{0}}{1+\exp\left[\kappa(\tilde{\theta}-\theta_{\rm T})\right]},
\end{align}
where we choose $\kappa=50$ \citep[see][]{Parfrey2013}. The velocity field is recentered at a colatitude $\theta_{\rm C}$ in the simulation coordinates $\left(r,\theta,\phi\right)$ by applying a rotation $\mathbf{R}_\alpha$ \citep[Section 5.1.2][]{Mahlmann2020b} as $\mathbf{v}_{\rm T}=\mathbf{R}_{-\alpha}\cdot\mathbf{v}\left(\mathbf{\tilde{r}}\right)$, where $\mathbf{\tilde{r}}=\mathbf{R}_{\alpha}\cdot\mathbf{r}$, and $\mathbf{r}$ is the coordinate vector in the simulation domain. We use the same boundary condition for 3D simulations and 2D axisymmetric models; in 2D we position the simulation slice at the center of the twisting disk and, thus, generate a double-sheared profile. The perfect conductor boundary (Appendix~\ref{sec:perfectconductorboundary}) only requires the specification of the electric field parallel to the surface, expressed as $E^\theta$ and $E^\phi$. We assume that the magnetic field $B^r$ is frozen into the stellar surface and does not change during the simulated time \citep[see][]{Mahlmann2019}.

\section{Results}
\label{sec:results}
We present the results from various FFE simulations for $\omega_0=1/25\times c/R_*$, where $T_0 = 2\pi/\omega_0$ is the rotational period of the twisting disk. We begin
by outlining our results for axisymmetric (2D) models in Section~\ref{sec:axisymmetric}, before 
generalizing them to 3D in Section~\ref{sec:openversuslocal}.

\begin{figure*}
 \centering
 \includegraphics[width=0.74\textwidth]{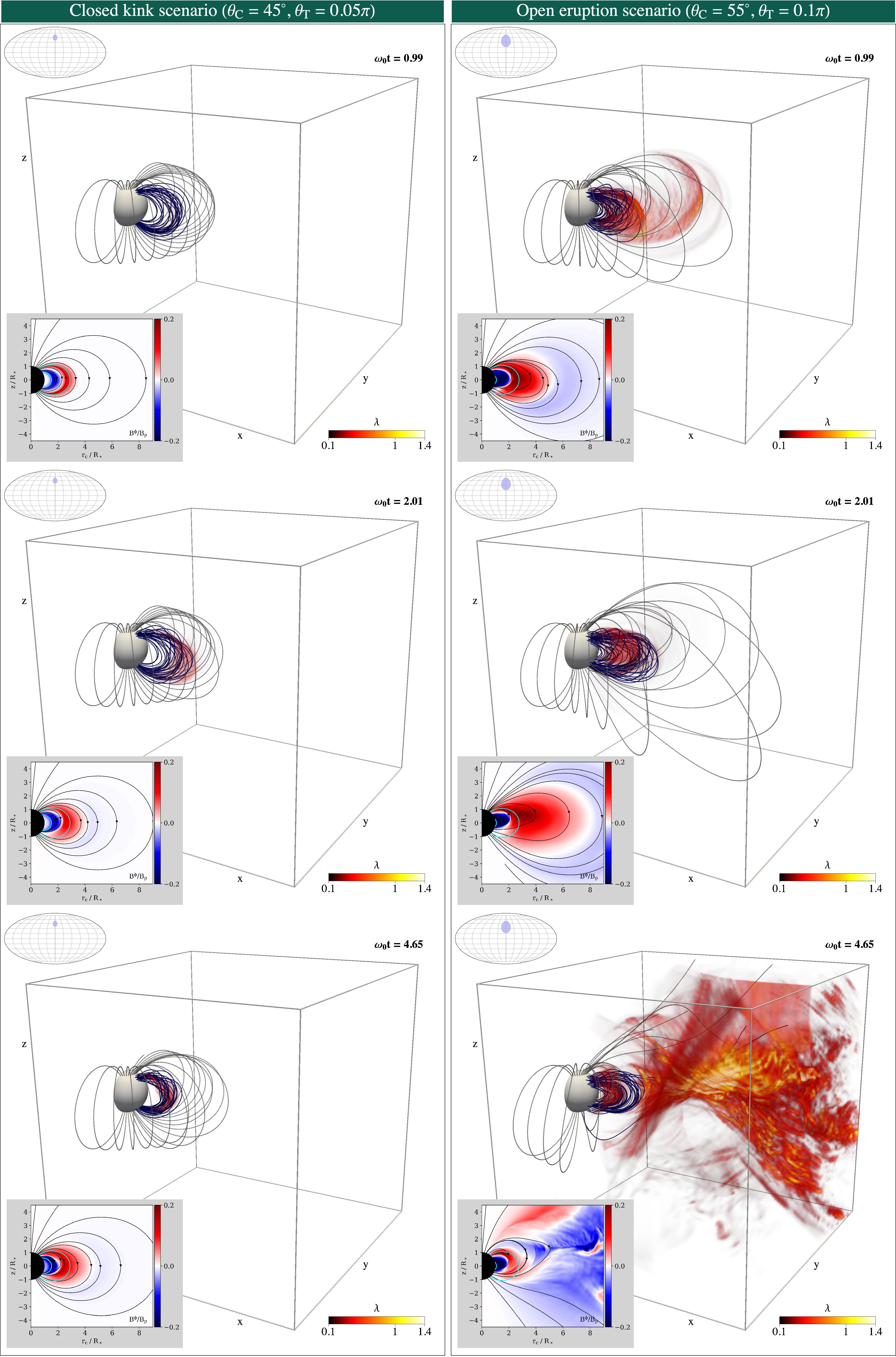}
 \caption{Three-dimensional visualizations of different flux tube evolution scenarios. We present different stages of the 3D eruption dynamics for varying geometries of the twisting region. Volume-filling colors denote the conserved parallel current $\lambda=\mathbf{j}_\parallel/|\mathbf{B}|$. Inset plots represent poloidal (field lines) and toroidal (color) magnetic fields in a meridional slice centered on the induced twist. In the closed kink scenario (left), the magnetic pressure outside of the twisted flux bundle prevents an eruption on global scales. Currents build up in the twisted flux tube, and energy is released repeatedly in helical, kink-like instabilities. The open eruption scenario (right) creates extended flux ropes that can open into large-scale current sheets.}
\label{fig:3DCompare}
\end{figure*}

\subsection{Axisymmetric (2D) models: Extended current sheets versus coronal flux ejections}
\label{sec:axisymmetric}

\emph{Lateral} eruptions that push an over-twisted flux tube through the external magnetic field and eventually open up the magnetosphere have been modeled in 2D by \citet[][]{Parfrey2012,Parfrey2013}. Appendix~\ref{app:2Deruptions} analyses axisymmetric eruption scenarios (set up according to Figure~\ref{fig:Figure1}, panel c). Here, we briefly summarize novel results.

In all our simulations, axisymmetric magnetospheres erupt when the footpoint displacement angle is larger than a critical value $\psi_{\rm crit}$ \citep{Parfrey2013}.
By analyzing 32 eruptions in our simulations, we find the critical angle (and standard errors) to be approximately
\begin{align}
\psi_{\rm crit}/(2\pi)=\left(0.26\pm 0.02\right)+\left(0.86\pm 0.04\right)\times \theta_{\rm E},
\label{eq:2Dcritical}
\end{align}
for axisymmetric flux bundles located at the latitude $\theta_{\rm E}$. For typical values of $25^\circ<\theta_{\rm E}<40^\circ$, Equation~\ref{eq:2Dcritical} yields $1.3\pi<\psi_{\rm crit}<1.7\pi$, or roughly $\psi_{\rm crit}\gtrsim \pi$. During each eruption, around $25\%$ of the injected twist energy is dissipated and potentially available for powering X-ray emission (see also Figure~\ref{fig:2DEruptions}). We measure dissipation as an immediate loss of magnetic energy by enforcing the FFE constraints $\mathbf{E}\cdot\mathbf{B}=0$ and $E^2-B^2<0$. Prior to the development of lateral instabilities, axisymmetric configurations with twist buried deep within the dipole magnetosphere can accumulate significant excess energy above the initial dipole field energy, $e_{\rm twist}=e_{\rm total}-e_{\rm dip}$. In our simulations, appropriate boundary shears can produce configurations with energies up to $e_{\rm twist}/e_{\rm dip}\approx 0.6$. The dynamics of magnetic field lines during eruptions of interior flux bundles pushing against outer flux layers were coined `magnetic detonation' in \citet[][see also \citealt{Cowley1997}]{Parfrey2013}. Especially when flux bundles are confined by layers of opposite twist, topologies similar to the erupting flux rope model for the solar photosphere \citep{Forbes1995} can develop. 

Following \citet{Gerrard2004} and \citet{Torok2010}, we suggest that the amount of magnetic flux in a twisted tube is the key factor for determining the eruption dynamics. We compare the enveloping dipolar field energy surrounding a flux tube \emph{without twist}, $e_{\rm env}$, to the magnetic energy $e_{\rm fb}$ of the flux bundle itself. For a dipolar equilibrium magnetosphere, we can employ the dipolar coordinates $\chi=\cos\theta/r^2$ and $\eta=r/\sin^2\theta$. They follow magnetic field lines parameterized by footpoint latitude $\sin^2\theta_0=R_*/r_{\rm eq}$ and equatorial extension $r_{\rm eq} = \eta$. With a suitable inverse transformation \citep{Swisdak2006} we can then define the magnetic energy $e_{\rm dip}(r_{\rm eq}^-,r_{\rm eq}^+)$ enclosed by two field lines with equatorial extensions $r_{\rm eq}^-$ and $r_{\rm eq}^+$. For a flux bundle centered at colatitude $\theta_{\rm C}$ with extension radius $\theta_{\rm T}$, we use 
\begin{align}
 e_{\rm env}(\theta_{\rm C},\theta_{\rm T}) &= e_{\rm dip}(R_*,\infty)-e_{\rm dip}(R_*,r_{\rm eq}^{\theta_{\rm C}-\theta_{\rm T}}),\\
 e_{\rm fb}(\theta_{\rm C},\theta_{\rm T}) &= e_{\rm dip}(R_*,r_{\rm eq}^{\theta_{\rm C}-\theta_{\rm T}})-e_{\rm dip}(R_*,r_{\rm eq}^{\theta_{\rm C}+\theta_{\rm T}}).
\end{align}
One expects a disruption of the equilibrium magnetospheric energy balance by twisting the field line footpoints for $e_{\rm fb}\approx e_{\rm env}$, or
\begin{align}
\begin{split}
 \mathcal{E}\equiv e_{\rm env}(\theta_{\rm C},\theta_{\rm T})-e_{\rm fb}(\theta_{\rm C},\theta_{\rm T})\approx 0.
 \label{eq:energybalance}
\end{split}
\end{align}
In Appendix~\ref{app:2Deruptions} (Figure~\ref{fig:2DEruptions}, left panel), we show that Equation~(\ref{eq:energybalance}) is a good approximate criterion for the onset of instabilities in the axisymmetric dipole magnetosphere and the available excess energy $e_{\rm twist}$.

\subsection{Non-axisymmetric (3D) models: Opened magnetospheres versus confined eruptions}
\label{sec:openversuslocal}

While axisymmetric twists always erupt when reaching a certain critical twist angle, the 3D evolution allows for significantly different dynamics. In this work, we identify the development of 3D \emph{helical (kink)} instabilities, and we ask \emph{which magnetospheric conditions favor such scenarios as opposed to lateral eruptions.} To illustrate the different possibilities qualitatively, we conducted two $\left[N_r,N_\theta,N_\phi\right]=\left[64,200,400\right]$ simulations with varying twisting disk centers at $\theta_{\rm C}\in\left[45^\circ,55^\circ\right]$ for extensions of $\theta_{\rm T}\in\left[0.05\pi,0.1\pi\right]$. 

In 2D, one can define a twist angle by measuring the helical footpoint displacement of magnetic field lines. In non-axisymmetric 3D magnetospheres, the measurement of $\psi$ is not straightforward. We use the projection of the conserved force-free current along the magnetic field to estimate the twist of flux tubes. This component of the current can be written in the form $\mathbf{j}_\parallel=\lambda\mathbf{B}$, with $\nabla\lambda\cdot\mathbf{B}=0$, hence, $\lambda$ constant along magnetic field lines. In a simplistic analogy to cylindrically symmetric force-free flux tubes, one can then use $\lambda$ to estimate the toroidal field component revolving around the flux tube with characteristic radius $r_0$ and guide field $B_\parallel$: 
\begin{align}
     B_T=\frac{I}{2\pi r_0}=\frac{r_0j_\parallel}{2}=\frac{r_0}{2}\lambda B_\parallel\;.
\end{align}

Figure~\ref{fig:3DCompare} identifies the two main evolution scenarios in 3D. The left column shows a helically unstable scenario developing a confined eruption. Twisted field lines close to the equator develop a helical instability at $\omega_0 t\approx 1$. In this scenario, a current sheet forms inside the twisted flux tube in the immediate magnetar vicinity. Here, magnetic reconnection occurs with a significant guide field (see Section~\ref{sec:longterm}). The right column of Figure~\ref{fig:3DCompare} shows a laterally unstable scenario. Field lines reach a critical shear and quickly open up the weaker dipole field of the extended magnetosphere in a large-scale eruption. Very similarly to the axisymmetric case (Figure~\ref{fig:2DEvol}, panel c) a current sheet forms for a short time in the tail of the magnetic ejecta expelled during the instability. We note that such current sheets are similar to the ones found in the wake of nonlinear Alfvén waves opening up the magnetosphere \citep[`pancakes', see][]{yuan2020,yuan2021}. In both confined and global eruptions, reconnection during instabilities rearranges the field topology by attaching field lines to different footpoints on the magnetar surface. Thus, relaxed magnetospheres after an eruption are non-axisymmetric and more complex than the initial dipolar configurations.

\newpage

\subsubsection{Eruption dynamics for different twisting geometries}
\label{sec:hotspotcharacteristics}

\begin{figure}
 \centering
 \includegraphics[width=0.47\textwidth]{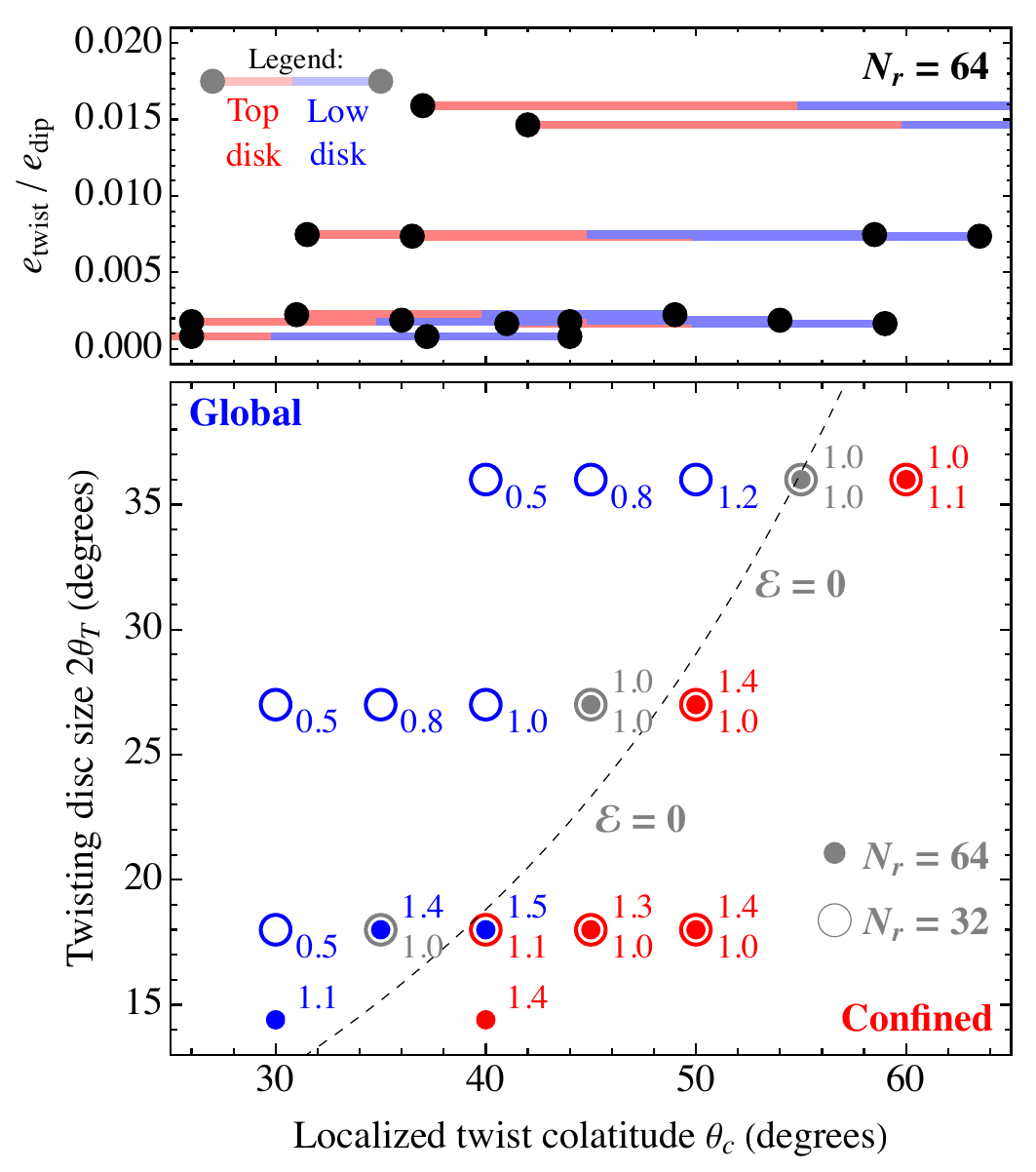}
 \vspace{-11pt}
 \caption{Parameter scan of the twisting disk colatitude $\theta_{\rm C}$ and extension $\theta_{\rm T}$ for 3D non-axisymmetric twist profiles (Section~\ref{sec:openversuslocal}) during a time interval of $\omega_0 t=2.1$. Top panel: Spatial extent of the twisting region in a meridional slice centered on the induced twist and the excess energy during the first eruption. Bottom panel: For different resolutions (rings: low; solid dots: high) we indicate if the magnetosphere erupts globally (blue) or dissipates during a confined, kink-like episode (red). Intermediate cases at the transition are colored in gray. Numbers adjacent to the data points denote at what time $\omega_0 t$ the respective dissipative event occurred, with lowercase scripts denoting low resolution and uppercase scripts for high resolution. Differences in the eruption time for higher resolutions can be explained by the efficient reduction of numerical diffusivity (see Appendix~\ref{app:calibationbenchmarks}). The dashed line corresponds to the approximate criterion given by the energy balance presented in Equation (\ref{eq:energybalance}).}
\label{fig:Figure3}
\end{figure}

\begin{figure*}
 \centering
 \includegraphics[width=1\textwidth]{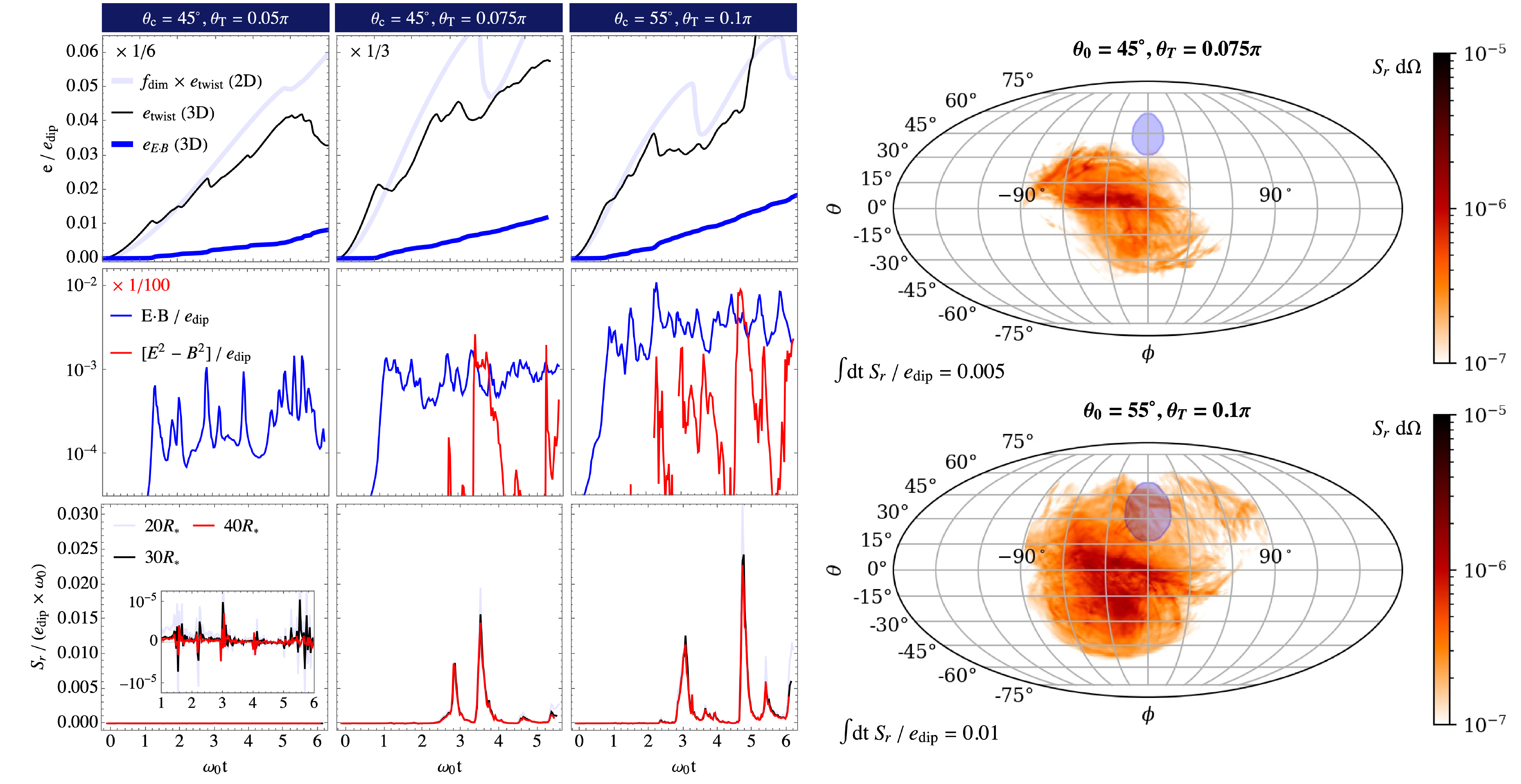}
 \vspace{-6pt}
 \caption{Magnetospheric energetics for selected long-term evolution scenarios of magnetospheric flux tubes ($\theta_{\rm C}\in\left[45^\circ,55^\circ\right]$, $\theta_T\in\left[0.05\pi,0.075\pi,0.1\pi\right]$). Left panel: We display the evolution of normalized energy measures $e/e_{\rm dip}$, namely the twist energy $e_{\rm twist}$, its 2D comparison weighted by $f_{\rm dim}$ (Equation~\ref{eq:dimrescaling}), and the energy $e_{E\cdot B}$ dissipated by FFE constraint conservation (top row). We also show dissipation measures for nonideal electric fields (middle row), and the outgoing Poynting flux at selected distances (bottom row; shifted in time to overlap). Right panel: Time-integrated Poynting flux through a surface at $40R_*$. The color maps show the angular distribution of the energy flow with respect to the twist injection site (shaded blue).}
\label{fig:LONGSIMS}
\end{figure*}
 
In Section~\ref{sec:openversuslocal} we identify two distinct 3D evolution scenarios, confined and global eruptions. In this section, we classify different twisting geometries by their eruption dynamics. Guided by the energy balance examined in Section~\ref{sec:axisymmetric} (Equation~\ref{eq:energybalance}), we conduct a parameter scan varying the magnetospheric energy balance, as to say, the twisting disk colatitude $\theta_{\rm C}$ and angular size $\theta_{\rm T}$ for different resolutions $N_r=\left[32,64\right]$ and $N_\theta=\left[100,200\right]$. Building up on our results of the previous section, we expect flux tubes located closer to the poles to be less confined ($\mathcal{E}<0$, global eruptions with open magnetospheres) than those that are deeply buried by the dipole field close to the equator ($\mathcal{E}>0$, confined eruption in a helical, kink-like instability).

For a set of 25 simulations and an evolution time of $\omega_0 t=2.1$, which focuses on the first episode of the development of the flux tube instability, we track the magnetospheric dynamics and energy content. We then identify for which parameters the magnetosphere opens up, as well as those for which it develops a helical instability with a closed topology and localized dissipation. 

Independent of whether simulations result in global eruptions leading to the open magnetosphere or a localized kink event, we find that the first instability  occurs at $\omega_0 t\approx 1$. Figure~\ref{fig:Figure3} shows the transition between global and confined eruptions for varying the location and size of the twisted region. For example, for a twist located at $\theta_{\rm C}=45^\circ$ with extent $\theta_{\rm T}=0.05\pi$, the twist energy is released in a kink event. Increasing the extent to $\theta_{\rm T}=0.075\pi$ or changing the center latitude to $\theta_{\rm C}=35^\circ$ induces eruptions with significant lateral extension or a global opening of the magnetosphere.

Confirming our expectations in Section~\ref{sec:openversuslocal}, our simulations show that flux tubes with footpoints closer to the equator and smaller areas are more likely to develop localized helical instabilities. Twisting regions of a larger extent and closer to the poles trigger global eruptions that open up the magnetosphere. We find that the $\mathcal{E}=0$ criterion on the energy balance derived in Equation~(\ref{eq:energybalance}) is indeed a viable threshold to distinguish global and confined eruptions (see gray line in Figure~\ref{fig:Figure3}).

\subsubsection{Long-term evolution of magnetospheric flux tubes}
\label{sec:longterm}

\begin{figure}
 \centering
 \includegraphics[width=0.48\textwidth]{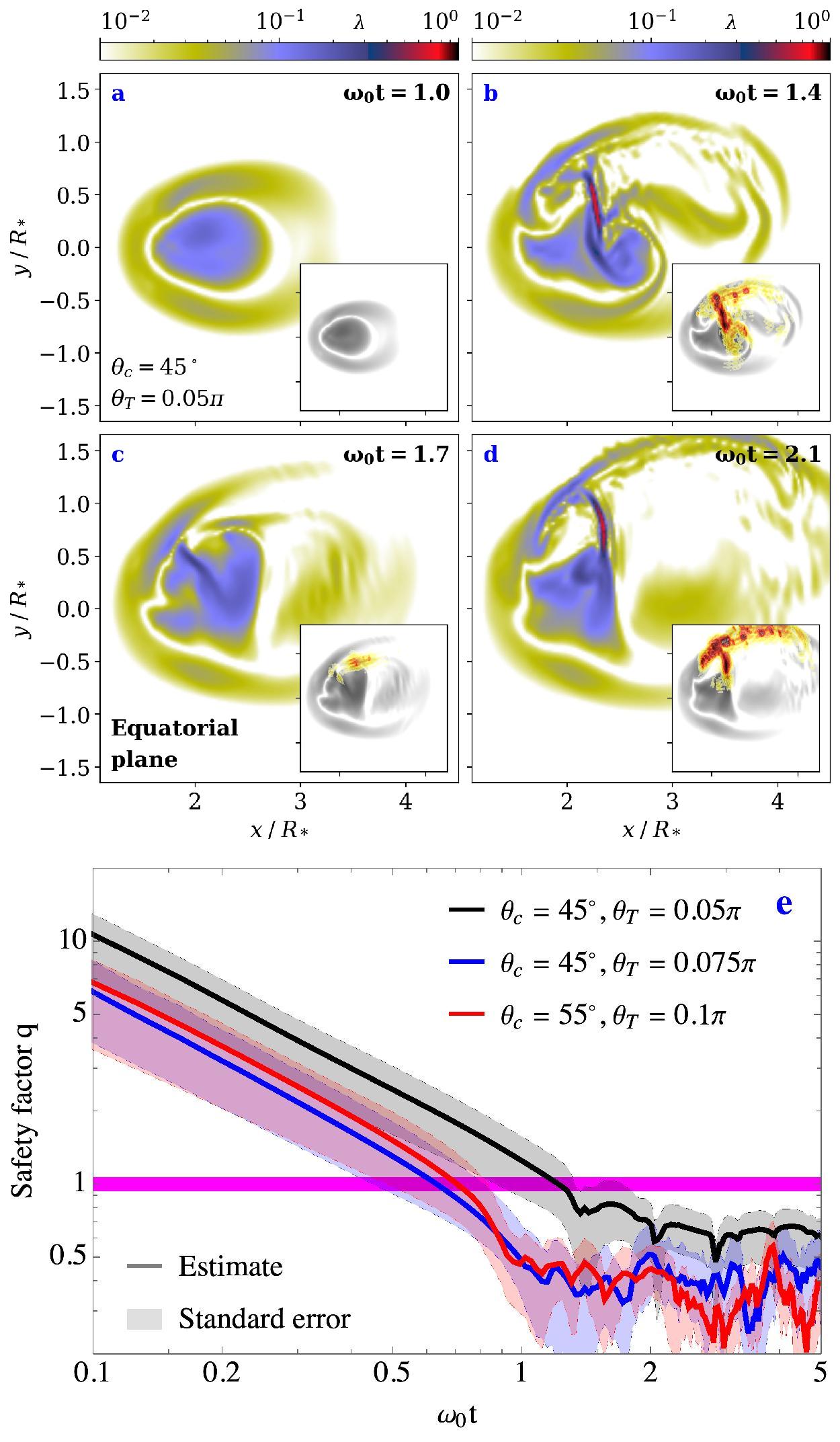}
 \vspace{-18pt}
 \caption{Build-up and release of magnetospheric twist during a helical instability and confined eruption. Panels (a) to (d) show cross sections of the parallel current $\lambda$ through the equatorial plane at different times. Insets indicate significant nonideal electric fields with $\mathbf{E}\cdot\mathbf{B}\neq 0$ in red color. Panel (e) displays an estimate of the flux tube safety factor. The thick magenta line denotes a common instability threshold ($q\approx 1$).}
 \vspace{-11pt}
\label{fig:TwistBuilt}
\end{figure}

In this section, we explore what happens if one continues twisting beyond the first instability episode to probe the eruption dynamics for larger excess energies. We simulate selected geometries of twisting regions for several characteristic (twisting) times. The chosen twist geometries are close to the $\mathcal{E}=0$ threshold between lateral and helical eruptions (dashed line in Figure~\ref{fig:Figure3}) and develop rich eruption dynamics at later times. We model the combinations $\left(\theta_{\rm C},\theta_T\right)\in\left[\left(45^\circ,0.05\pi\right),\left(45^\circ,0.075\pi\right),\left(55^\circ,0.1\pi\right)\right]$ for a duration of $\omega_0 t=6$ at a high resolution of $\left[N_r,N_\theta,N_\phi\right]=\left[64,200,400\right]$. Animations tracking the long-term evolution of the scenarios displayed in Figure~\ref{fig:3DCompare} can be found in the supplemental material \citep{SupplementaryMediaA,SupplementaryMediaB}.

In Figure~\ref{fig:LONGSIMS}, we show different measures of energy evolution. We find that the growth of long-term magnetospheric twist energy generally reproduces the rescaled rate of the respective 2D models (by weighting with the size of the non-axisymmetric twisted region; see Equation~\ref{eq:dimrescaling}). After the first eruptions, as described in Section~\ref{sec:openversuslocal}, further eruptions continue to develop over intervals of roughly $\omega_0\Delta t\approx1-1.4$. 

The magnetospheric instabilities are accompanied by the development of nonideal electric fields and a breakdown of FFE conditions ($\mathbf{E}\cdot\mathbf{B}\neq 0$). As is evident in the left panels of Figure~\ref{fig:LONGSIMS} (second row), significant $\mathbf{E}_\parallel$ develops roughly at the time of the first interruption for all models. Figure~\ref{fig:TwistBuilt} (panels a to d) shows cross sections of the conserved parallel current $\lambda$ through the equatorial plane for a confined eruption. During the helical instability (right column; panels b and d), narrow regions of strong currents form. Current sheets appear during reconnection with a significant guide field (Figure~\ref{fig:TwistBuilt}, panel b), about three to five times stronger than the reconnecting magnetic field, similar to the current sheets seen in studies of the kink instability \citep[see][Figure~2]{Davelaar2020}. These current layers also have localized regions of large $\mathbf{E}\cdot\mathbf{B}\neq 0$, displayed in red color in the respective inset plots. Between eruptions, currents are more uniformly distributed and nonideal electric fields are less pronounced (Figure~\ref{fig:TwistBuilt}, panel c). In contrast, the globally erupting configurations produce an intermittent development of electrically dominated regions ($E>B$), which are the indicators of the low guide field reconnection in the tail of the ejecta (see Figure~\ref{fig:LONGSIMS}, also similar to 2D models, Figure~\ref{fig:2DModels}). In our simplistic treatment of nonideal plasma regions, these violations are quickly dissipated by the FFE constraint enforcement. In reality, we expect these locations to be sites of efficient X-ray generation by radiative reconnection \citep[e.g.,][]{Beloborodov2021}.

In the two globally erupting setups, we also measure significant Poynting fluxes in the ejecta of the global reconnection event. $S^r$ denotes the surface-integrated radial Poynting flux through spheres at different radii. In Figure~\ref{fig:LONGSIMS}, we normalize $S^r$ to the twisting time scale $\omega_0$, and the dipole energy $e_{\rm dip}$ (see also discussion in Section~\ref{sec:discussion}). We show the distribution of the Poynting-flux ejecta (right panel of Figure~\ref{fig:LONGSIMS}), propagating outward. Up to $1\%$ of the initial dipole field energy can be carried away to the outer magnetosphere in the form of magnetically dominated ejecta and potentially power radio bursts in the outer magnetosphere (see discussion in Section~\ref{sec:discussion}). Between the two cases we explored, larger twisting regions buried deeper in the magnetosphere produce more energetic ejecta covering larger solid angles. Confined eruptions show fainter levels of energy outflows (see Figure~\ref{fig:LONGSIMS}), about four orders of magnitude lower than the global eruptions. The energy is carried by fast magnetosonic waves correlated with the kink episodes. These waves are injected with an amplitude of $\delta B/B_d\approx 10^{-3}$ at a distance of $2R_*$ to $3R_*$. We measure their wavelength in the range of $\lambda\approx 0.1 - 1.0 R_*$, which corresponds to a frequency of $10 - 100$kHz. As they propagate outward across magnetic field lines, in our force-free simulations, such waves develop electrically dominated regions, $E>B$, at distances between $60 R_*$ and $100 R_*$ and dissipate via FFE constraint enforcement. Similar wave parameters are considered in the magnetohydrodynamic limit by \citet[][see Section~7]{Beloborodov2022}, who showed that such waves will result in formation of strong shocks in the magnetosphere.

An important measure for predicting the onset of a magnetospheric instability is the safety factor, $q$. Commonly, $q\lesssim 1$ indicates that a flux tube is susceptible to instabilities \citep[see discussion in, e.g.,][]{longaretti2008}:
\begin{align}
q\equiv\frac{2\pi r_0}{L}\frac{B_\parallel}{B_T}=\frac{4\pi}{\lambda L}.
 \label{eq:safetydefinition}
\end{align}
Here, $B^T$ is the toroidal magnetic field in a cylindrical coordinate system revolving around $B_\parallel$, and $L$ is a characteristic length of the flux tube. In practice, the safety factor for a dipolar geometry can be nonuniform through the twisted flux tube. However, in the following, we evaluate it by making a few basic assumptions. First, we measure the average parallel current $\lambda$ within the (static) region initially enclosed by the flux tube during the linear growth of the twist. Second, we fix $r_0$ as the radius of a disk with the area of this region, and $L$ to the average initial length of the flux tube. Figure~\ref{fig:TwistBuilt} (panel e) shows the steady decline of the estimated safety factor prior to the first eruption of the magnetospheres, ultimately leading to an onset of instability ($q\lesssim 1$, magenta line) around $\omega_0 t \approx 1$. In all cases, $q$ saturates after the first energy release due to magnetospheric instabilities. We note that the models with global eruptions (blue and red lines in Figure~\ref{fig:TwistBuilt}) intermittently show small values of $q$, which is directly related to the formation of current sheets characterized by large values of $\lambda$.

\subsection{2D versus 3D comparison}

Axisymmetric twist perturbations of a dipolar magnetic field (set up, e.g., according to Figure~\ref{fig:Figure1}, panel c) always develop lateral instabilities for magnetospheric twists above the critical value $\psi_{\rm crit}$ (Equation~\ref{eq:2Dcritical}). Lateral eruptions open up the closed dipole magnetosphere in extended current sheets along the equator and electromagnetic ejecta (see Figures~\ref{fig:2DEvol} and~\ref{fig:2DModels}).

In 3D, the stability of twisted magnetospheric flux tubes needs to be evaluated in two aspects. First, the \emph{onset} of the instabilities. By estimating the respective safety factors in Equation~(\ref{eq:safetydefinition}), we take into account the flux tube geometry. For different geometries, the twisted magnetospheres become unstable for $q\lesssim 1$. Second, the specific \emph{manifestation} of the instability, or in other words, the position
in Figure~\ref{fig:Figure3} (see Section~\ref{sec:hotspotcharacteristics}). Eruptions of flux tubes with  $\mathcal{E}<0$, as to say confining magnetic energy below the energy of the flux bundle, extend to global scales producing large-scale current sheets and opening up the dipole magnetosphere \citep[global eruptions; see also][]{Carrasco:2019aas}. Flux tubes with $\mathcal{E}>0$ develop instabilities (confined eruptions) that happen in the inner magnetosphere and dissipate energy with helical, kink-like dynamics. 

The energy injected into the magnetosphere by the localized twist approximately scales with its area when compared to the axisymmetric case (Section~\ref{sec:axisymmetric}). The 3D/2D twisting region size ratio is
\begin{align}
 f_{\rm dim}\equiv\frac{A_{3D}}{A_{2D}}=\frac{1}{4}\times\frac{\sin\theta_T}{\sin\theta_{\rm C}}.
 \label{eq:dimrescaling}
\end{align}
Specifically, for the parameter scan presented in this section, we find $0.04<f_{\rm dim}<0.12$. Using the same footpoint displacement velocity, the critical safety factor $q\lesssim 1$ is reached at earlier absolute times in our 3D models (Section~\ref{sec:openversuslocal}) than $\psi_{\rm crit}$ is reached in their 2D counterparts (Section~\ref{sec:axisymmetric}, see also top left panels of Figure~\ref{fig:LONGSIMS}). This results in low twist energies during eruptions for most 3D models, a fact that becomes  especially obvious when comparing Figures~\ref{fig:Figure3} (top panel) and~\ref{fig:2DEruptions} (right panel). 

Flux tubes at higher latitudes along field lines with rapidly decaying field strength have especially low excess energies. Similarly to the axisymmetric twist evolution, only a fraction of the twist energy $e_{\rm twist}$ is dissipated during each eruption. In contrast to twists extending to large radii, one model with a deeply buried twist ($\theta_{\rm c}=55^\circ, \theta_{\rm T}=0.1\pi$, third column in the left panel of Figure~\ref{fig:LONGSIMS}) accumulates $e_{\rm twist}/e_{\rm dip}\approx 0.05$ after several eruptions. The excess energy that can be reached in twisted dipole magnetospheres with disc-like surface deformations was recently constrained by evaluating Grad-Rubin equilibrium configurations \citep{Stefanou2022}. For their most favorable twist geometry with a comparably large surface extension ($\theta_{\rm C}=45^\circ$, $\theta_{\rm T}\approx 0.15\pi$), \citet{Stefanou2022} find an equilibrium magnetosphere (i.e., a configuration that would still be stable against eruptions) with similarly low excess energy of $e_{\rm twist}/e_{\rm dip}\approx 0.06$. We further discuss the energy accumulation and release during observed magnetar bursts in Section~\ref{sec:discussion}.

\section{Discussion}
\label{sec:discussion}

\begin{figure}
 \centering
 \includegraphics[width=0.47\textwidth]{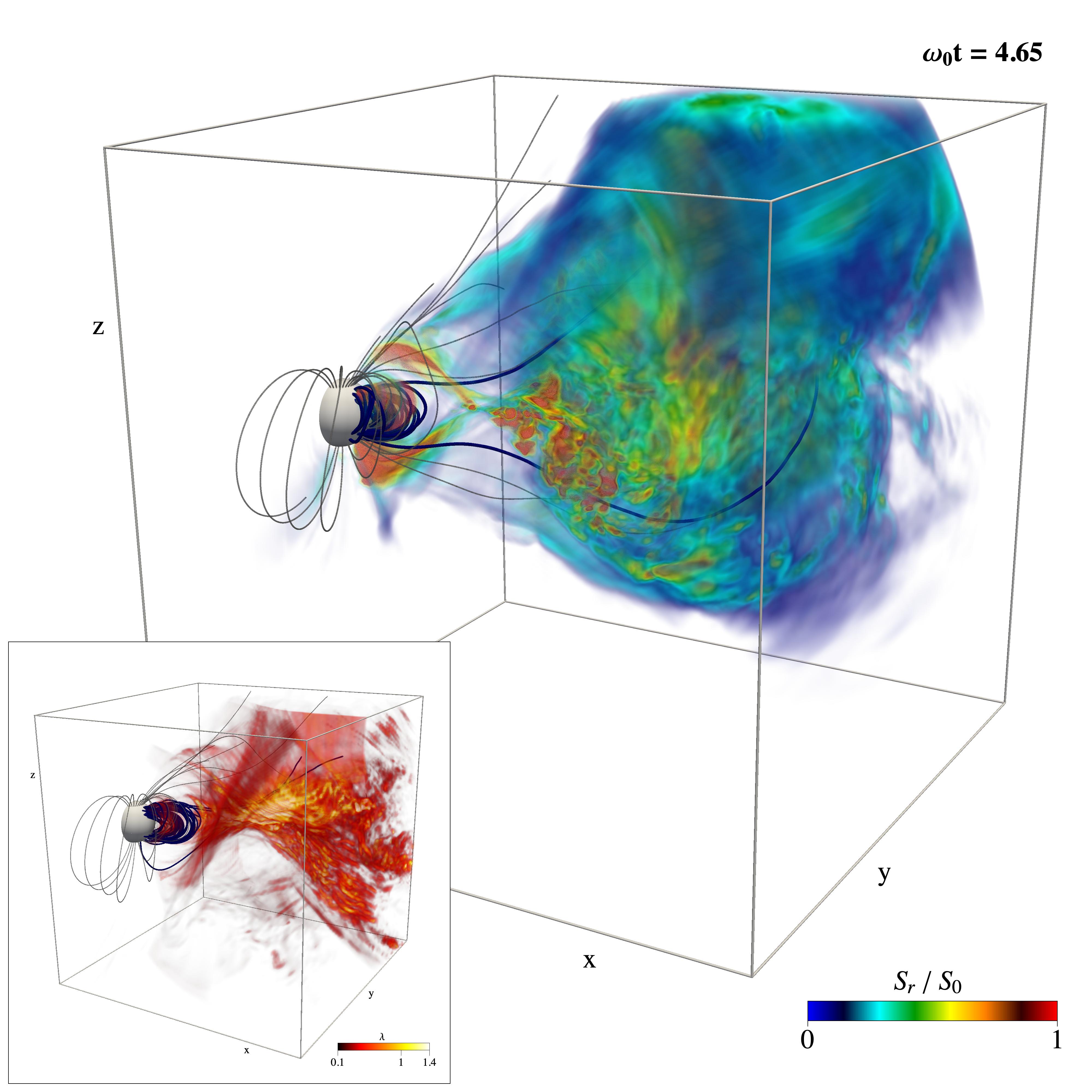}
 \vspace{-11pt}
 \caption{Ejection of Poynting flux along the radial direction during an eruption of the $\theta_{\rm c}=55^\circ$, and $\theta_{\rm T}=0.1\pi$ model. During a large-scale reconnection episode, as captured by the parallel current $\lambda$ in the inset and Figure~\ref{fig:3DCompare} (bottom panel), a blob of radially outward-flowing energy is released from the bursting site.}
\label{fig:Figure5}
\end{figure}

In this Letter, we study how twisted magnetic fields resulting from deformations of the magnetar crust induce magnetospheric instabilities. In the following, the stellar radius $R_*$ and magnetic field $B_*$ are scaled to typical reference cases in the magnetar population.

\subsection{Implications for X-ray bursts and giant flares}

In principle, the magnetic energy dissipated rapidly by enforcing the FFE constraints during magnetospheric instabilities can be converted to X-ray emission. Assuming $e_{\rm diss}/e_{\rm twist}\approx 0.25$ (Figure~\ref{fig:2DEruptions}, right panel) and $e_{\rm diss}\approx 10^{46}\text{erg}$ for giant flare events like the initial spike of SGR 1806-20 \citep{Palmer2005} sets a limit of $e_{\rm twist}\gtrsim 4\times 10^{46}\text{erg}\approx 0.13\times e_{\rm dip}$ as a requirement for the brightest observed events. Some axisymmetric models can store magnetic energies in excess of the dipolar field of up to $e_{\rm twist}\approx 0.6\times e_{\rm dip}$ (see Figure~\ref{fig:2DEruptions}), where
\begin{align}
 e_{\rm dip}\approx 3\times 10^{47}\left(\frac{B_*}{10^{15}\text{G}}\right)^2\left(\frac{R_*}{10\text{km}}\right)^3\text{erg}.
\end{align}
Several of them can easily reach the giant flare limit, as we show in Figure~\ref{fig:2DEruptions} \citep[see also the discussion by][Section 7.4]{Parfrey2013}. Additional dissipation can occur during secondary events triggered by instabilities in the inner magnetosphere, most notably shocks or reconnection in the outer magnetosphere (see Section~\ref{sec:frbimplications}).

Our models show that in 3D, there is no straightforward accumulation of giant-flare-type energies in flux tubes deeply buried in the magnetar magnetosphere. Prior to the first eruption, the excess energy of the twisted configurations is usually of the order of one percent of $e_{\rm dip}$ (see Figure~\ref{fig:Figure3}). Thus, the dissipated energy is of the order of $0.001\times e_{\rm dip}$ or less (assuming a conversion efficiency\footnote{While this upper limit is based on the analysis of multiple axisymmetric eruptions, our 3D results show a similar dissipation efficiency on a limited set of both global and confined eruptions. We note that \citet{Parfrey2013} use a lower value of $e_{\rm diss}/e_{\rm twist}\approx 0.1$. However, such dissipation estimates have some uncertainty (see scatter in the right panel of Figure~\ref{fig:2DEruptions}) and should only be used as upper limits. Therefore, the results obtained in this work do not contradict the results by \citet{Parfrey2013}.} of $e_{\rm diss}/e_{\rm twist}\lesssim 0.25$), which falls into the range of typical X-ray outburst activity \citep[$10^{41}-10^{43}$erg, see Section 2.2.3 in][]{Esposito2020}. Still, we find that scenarios with continuing twists allow for consecutive eruptions of 3D flux structures and increase the overall magnetospheric energy in between them. It is impossible for highly energetic flares to occur in isolation due to magnetospheric instabilities induced by narrow twisting regions. However, the most luminous events could follow less energetic precursor flares emerging from the kink-like instabilities of the continuously twisting flux bundle. 

\subsection{Implications for fast radio bursts (FRBs)}
\label{sec:frbimplications}

During 3D eruption events, part of the energy is ejected in packets of Poynting flux that propagate away from the central object (see Figures~\ref{fig:LONGSIMS} and~\ref{fig:Figure5}). For global eruptions, the luminosity associated to the outgoing Poynting flux shown in Figures~\ref{fig:LONGSIMS} and~\ref{fig:Figure5} is $L_S\approx 5.7\times 10^{48}\text{erg}/\text{s}$. Here, we used $S^r/(\omega_0\times e_{\rm dip})\approx 0.015$ from Figures~\ref{fig:LONGSIMS} and we used the simulation-specific value of
$\omega_0=1/25\times c/R_*$. The ejecta produced in global eruptions have a significant fast magnetosonic component. Similar to the `pancake' structures modeled by \citet{yuan2020}, \citet{Yuan2022} in the regime of fast shearing and \citet{Sharma2023} for slowly sheared magnetospheres, they propagate outward as nonlinear structures. Outgoing electromagnetic pulses of luminosities in this range can power coherent GHz emission in the outer magnetosphere, e.g., via synchrotron maser radiation \citep{Lyubarsky2014,Ghisellini2016,Beloborodov2017,Metzger2019,Plotnikov2019,Beloborodov2020,Margalit2020,Sironi2021} or mode conversion \citep{Thompson2022} at relativistic magnetized shocks in the magnetar wind, or via forced magnetic reconnection in the compressed current sheet of the magnetosphere \citep{lyubarsky2020,Mahlmann2022}. Our globally erupting configurations are thus capable of powering both the X-ray bursts and FRB-like transients. Confined, kink-like events generate fainter outflows of low-frequency fast magnetosonic waves of luminosity $L_S\lesssim 1.8\times 10^{45}\text{erg}/\text{s}$, using $S^r/(\omega_0\times e_{\rm dip})\lesssim 5\times 10^{-6}$ (see Figure~\ref{fig:LONGSIMS}). These waves become nonlinear within the magnetar light cylinder when approaching a state with $E\approx B$. They are capable of dissipating electromagnetic energy by developing shocks and heating the plasma \citep{Beloborodov2022,Chen2022,Levinson2022}. It is unlikely that these modes provide sufficient energy at large distances for powering reconnection-mediated FRBs \citep{lyubarsky2020,Mahlmann2022}. However, shocks induced by the breakdown of FFE could be an alternative channel for the generation of fainter X-ray bursts \citep{Beloborodov2022}.

\subsection{Limitations}

Our models shear the footpoints of magnetic field lines with an angular velocity $\omega_0=\chi_\omega \times c/R_*$, where $\chi_\omega$ is a dimensionless constant. In a realistic magnetosphere, we would expect the magnetar light cylinder to be at a large distance from the twisted flux tube. Therefore, we do not include the magnetar rotation in our models. Moreover, the footpoint rotation rate governed by slow processes in the magnetar crust should be very slow, $\chi_\omega \ll 1$. We choose it to be as slow as permissible via numerical diffusion (see Appendix~\ref{app:calibationbenchmarks}), with a default choice of $\chi_\omega=1/25$. To probe the validity of results in Section~\ref{sec:results} for slower twisting of field lines, we verified the development of selected 3D magnetospheric instabilities for $\chi_\omega=1/50$, and $\chi_\omega=1/100$. While numerical diffusivity notably affects the excess energy for slowest rotation rates (see Appendix~\ref{app:calibationbenchmarks}), the magnetospheres erupt at roughly $\omega_0 t\approx 1$, confirming the results of reference cases with a faster twist of $\chi_\omega=1/25$. Neglecting the magnetar rotation prevents us from measuring the impact of 3D instabilities on the magnetar spin-down \citep{Parfrey2012,Parfrey2013}, which is important for interpreting spin-down anti-glitches \citep[see, e.g.,][for a recent event in SGR J1935+2154 triggering increased radio activity]{Younes2022a}. We defer the study of rotating magnetospheres to future work.

Besides the twisting time scales, the twisting geometry we employ in this work is equally idealized: in 2D, we use twisting axisymmetric rings, and in 3D we model twisting discs on the magnetar surface. In reality, the region of twist injection may be complex of field line footpoints moving in opposite directions along fault-like structures \citep[e.g., Figure 2 in][]{Thompson2017}, for example, the induced field line deformations may be more ellipsoidal, or a ring-like structure may extend only over parts of the stellar surface. In these cases, a significant amount of energy close to the $e_{\rm twist}\gtrsim 4\times 10^{46}\approx 0.13\times e_{\rm dip}$ threshold could be stored in the magnetosphere, while the complex twist geometry may still trigger episodes of precursor flares. Constraining the exact profile and dynamics of field line motion near the magnetar surface in future works will require a combined understanding of crustal processes and their connection to the outer magnetosphere.

Finally, we note that the crude treatment of dissipation in FFE schemes dilutes correct estimates of the dissipative time scales for both the lateral (global) and helical (local) instabilities identified in our simulations \citep[see][]{Mahlmann2020c}. Our mostly kink-unstable reference case ($\theta_{\rm c}=45^\circ, \theta_{\rm T}=0.05\pi$, first column in the left panel of Figure~\ref{fig:LONGSIMS}) shows brief periods of dissipation followed by a further increase in excess energy. Contrasting this, global eruptions (accompanied by regions with $E>B$, see second and third columns in the left panel of Figure~\ref{fig:LONGSIMS}) show longer-lasting dissipation episodes. In principle, the eruption time scale is $\Delta t\approx L/v_{\rm diss}$ \citep[see Equation~25 in][]{Parfrey2013}, where $v_{\rm diss}$ is the dissipation rate. In our specific application, it can vary in two ways. First, the length $L$ of the affected field lines is longer in global eruptions extending to larger radii. Second, the dissipation rate can be different in the large-scale current sheets of global eruptions (determined by the reconnection rate in a single sheet, $v_{\rm diss}\approx v_{\rm rec}$) and the small-scale reconnection regions in confined kink events. It is possible that the dissipation timescale during kink events occurring locally and deep inside the magnetosphere can be distinguished from global events that open up the magnetar magnetosphere. The points identified in this section have implications for X-ray burst energies and durations, and we will revisit these issues in future work incorporating realistic dissipation physics.

\section{Conclusion}
\label{sec:conclusion}

Our simulations unveil a new type of magnetospheric instability of twisted flux tubes occurring well within the magnetar magnetosphere. Besides the well-established lateral eruption (global eruption) of twisted magnetospheres \citep{Parfrey2013}, 3D structures can develop helical, kink-like dynamics and dissipate energy locally (confined eruptions). During such events, a fraction of the twist energy, likely $e_{\rm diss}/e_{\rm twist}\lesssim 0.25$, is dissipated and available to power X-ray outbursts. In addition to providing this energy reservoir, global eruptions open up the magnetosphere in extended current sheets and outgoing energy flows. They can provide the necessary feedback on the outer magnetosphere required by so-called far-away FRB models that inject high-frequency radio waves by shocks or magnetic reconnection. While 3D twist structures driven by narrow twisting regions on the magnetar surface likely do not reach giant flare energy output, their rapid bursting activity could explain the most common X-ray outbursts. Powerful eruptions with significant energy built up deeply buried in the dipole magnetosphere will be accompanied by less energetic outbursts due to the intermittent instabilities of a continuously twisting flux tube. Complex shearing geometries would likely allow for wider ranges in burst energy and rich combinations of bursting activity. The models described in this work are one ingredient of global scenarios of transient magnetar activity, where instabilities close to the star shine bright in the X-ray band and seed multiwavelength phenomena in the extended magnetosphere.

\section*{Acknowledgments}

We are grateful for the helpful suggestions by the referee, especially regarding the analysis of outgoing fast modes from confined eruptions. The authors thank A. Beloborodov, P. Cerdá-Durán, F. Bacchini, M. Gabler, H. Hakobyan, Y. Lyubarky, K. Parfrey, E. Novoselov, A. Spitkovsky, P. Stefanou, C. Thompson, and Z. Wadiasingh for useful discussions. We thank M. Elmassry for engaging in the Princeton Writing Program and giving comments to improve our manuscript. 

We are grateful for the funding provided through NASA grant 80NSSC18K1099. AAP and JFM acknowledge support by the National Science Foundation under grant No. AST-1909458. This research was facilitated by the Multimessenger Plasma Physics Center (MPPC), NSF grant PHY-2206607. VM is supported by the Exascale Computing Project (17-SC-20-SC), a collaborative effort of the U.S. Department of Energy (DOE) Office of Science and the National Nuclear Security Administration. Work at Oak Ridge National Laboratory is supported under contract DE-AC05-00OR22725 with the U.S. Department of Energy. Support for this work was provided by NASA through the NASA Hubble Fellowship grant HST-HF2-51518.001-A awarded by the Space Telescope Science Institute, which is operated by the Association of Universities for Research in Astronomy, Incorporated, under NASA contract NAS5-26555. ERM gratefully acknowledges support as the John A. Wheeler Fellow at the Princeton Center for Theoretical Science, the Princeton Gravity Initiative, and the Institute for Advanced Study. 

This research is part of the \textit{Frontera} \citep{Stanzione2020} computing project at the Texas Advanced Computing Center (LRAC-AST21006). \textit{Frontera} is made possible by the National Science Foundation award OAC-1818253. The presented numerical simulations were further enabled by the \textit{MareNostrum} supercomputer (Red Española de Supercomputación, AECT-2021-1-0006), and by the VSC (Flemish Supercomputer Center), funded by the Research Foundation Flanders (FWO) and the Flemish Government -- department EWI. ERM acknowledges support through the Extreme Science and Engineering Discovery Environment \citep[XSEDE,][]{Towns2014} through Expanse at SDSC and Bridges-2 at PSC through allocations PHY210053 and PHY210074. The authors further acknowledge supported by Princeton Research Computing, a consortium of groups including the Princeton Institute for Computational Science and Engineering (PICSciE) and the Office of Information Technology's High-Performance Computing Center and Visualization Laboratory at Princeton University.

\bibliography{literature.bib}

\appendix

\section{Eruptions of axisymmetrically twisted magnetospheres}
\label{app:2Deruptions}

\begin{figure*}
 \centering
 \includegraphics[width=0.99\textwidth]{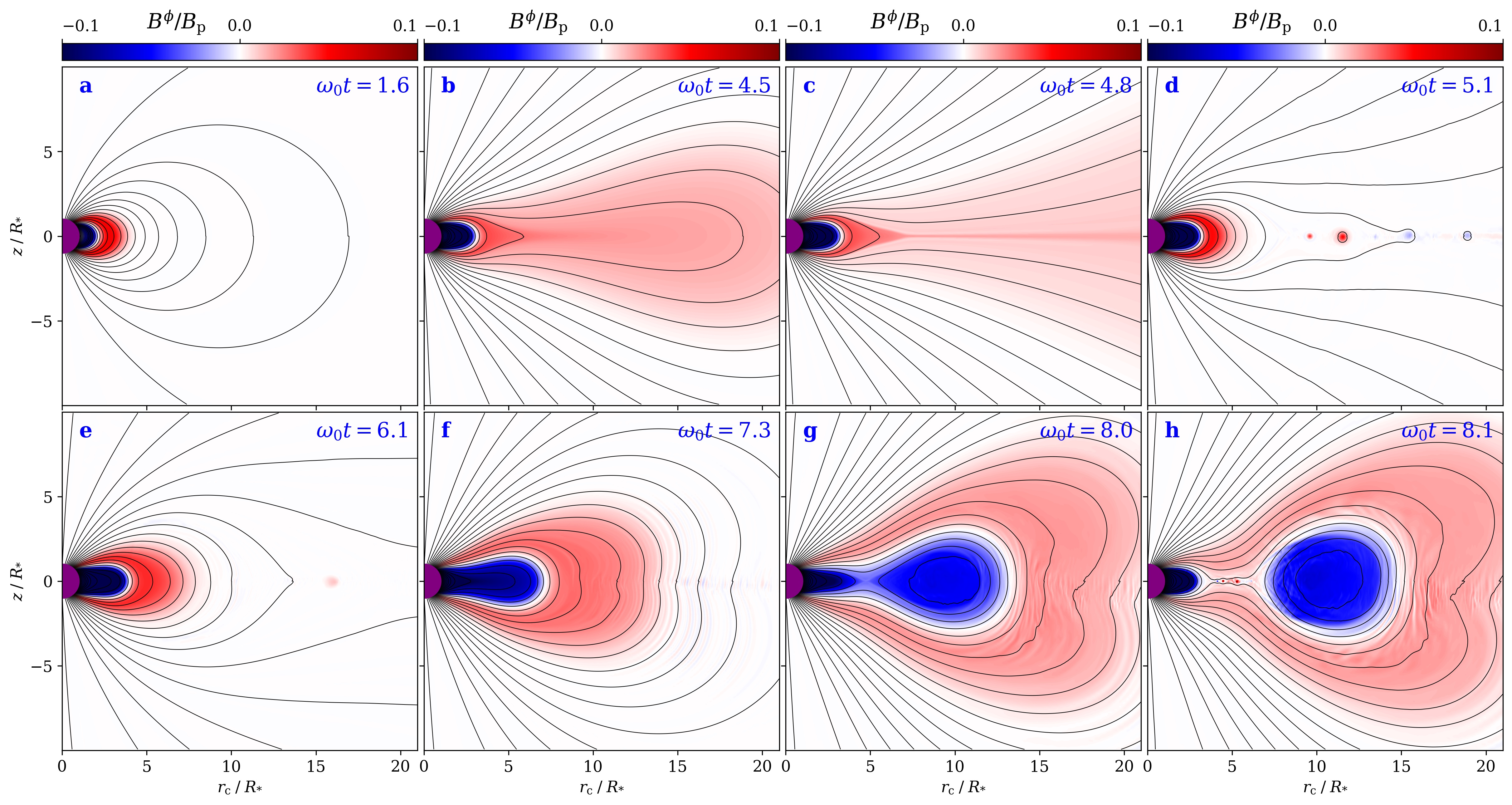}
 \vspace{-6pt}
 \caption{Eruption stages of an axisymmetric model with $\theta_{\rm C}=45^\circ$ and $\theta_{\rm T}=0.05\pi$. We present contours of the magnetic potential; colors denote the toroidal magnetic field normalized by the polar field strength $B_p\equiv B_d(\theta=0)$. The different stages are labeled as follows: (a) the initial twisting phase of the dipolar topology. (b)-(d) Eruption of the outer layer of twisted magnetic fields in an elongated current sheet, reconnecting with magnetic island mergers. (e) Relaxation to the dipolar geometry after expelling parts of the outer twist layer. (f)-(h) Eruption of the inner twist layer generating an electromagnetic ejecta. An animated version is provided in the supplemental material \citep{SupplementaryMediaC}.}
\label{fig:2DEvol}
\end{figure*}

\begin{figure}
 \centering
 \includegraphics[width=0.9\textwidth]{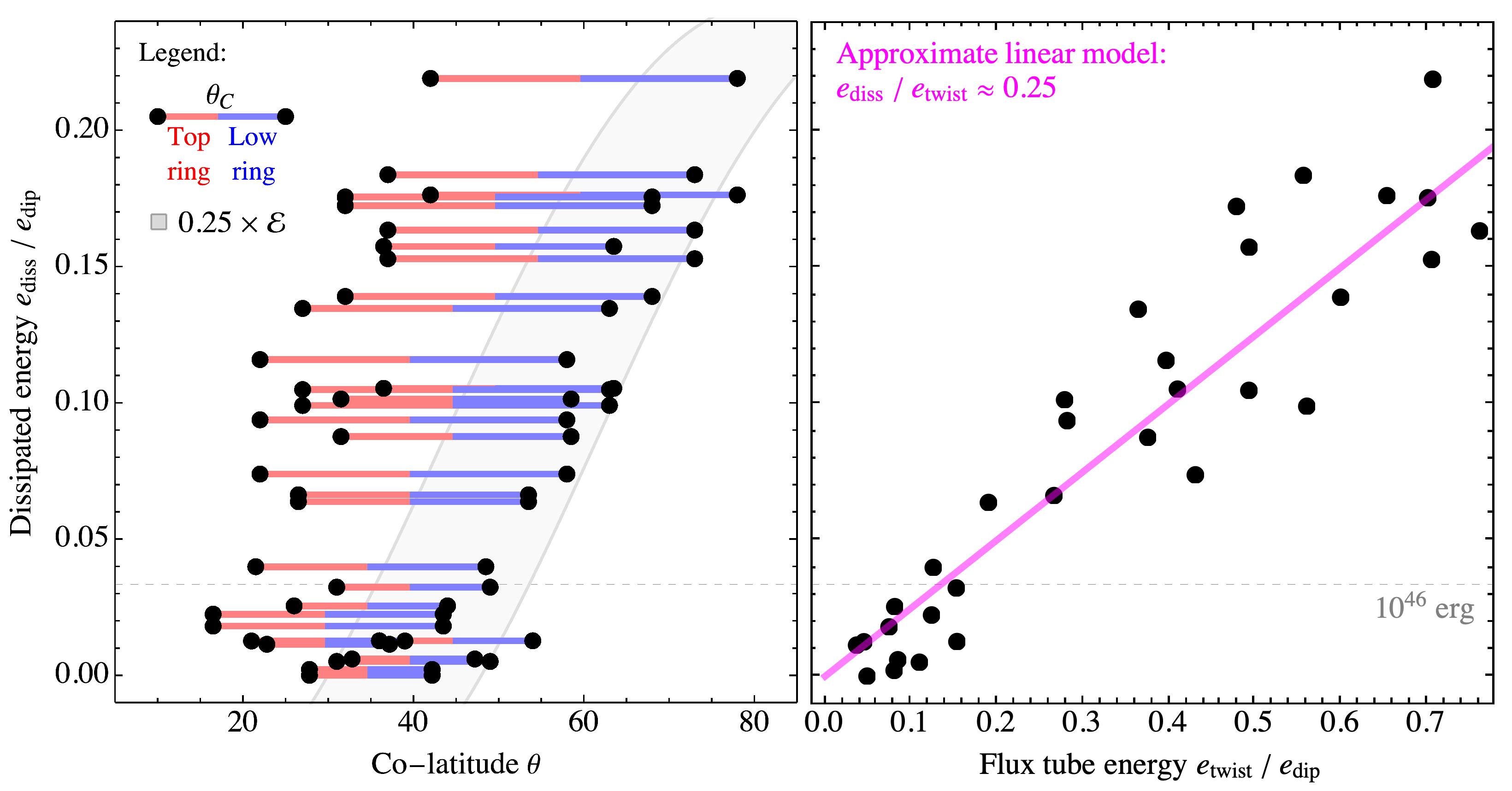}
 \vspace{-5pt}
 \caption{Energetics of 32 axisymmetric erupting magnetospheres obtained from simulating the setup described in Section~\ref{sec:axisymmetric}. \textit{Right:} Energy dissipated in an isolated eruption event versus total energy injected by twisting the magnetospheres (both normalized to the total energy of the equilibrium dipole magnetosphere). The magenta line denotes an approximate linear model for the dissipation with $e_{\rm diss}/e_{\rm twist}=0.25$. \textit{Left:} Dissipated energy as a function of the twist extension, denoted by a bar along its angular extension. The gray shaded region indicates the amount of dissipated energy expected when assuming that the magnetosphere is twisted up to its equilibrium threshold (as defined in Equation~\ref{eq:energybalance}). We note that the analyzed eruptions are mainly due to the instability of deeply buried magnetic twists (indicated in blue color).}
\label{fig:2DEruptions}
\end{figure}

\begin{figure}
 \centering
 \includegraphics[width=0.98\textwidth]{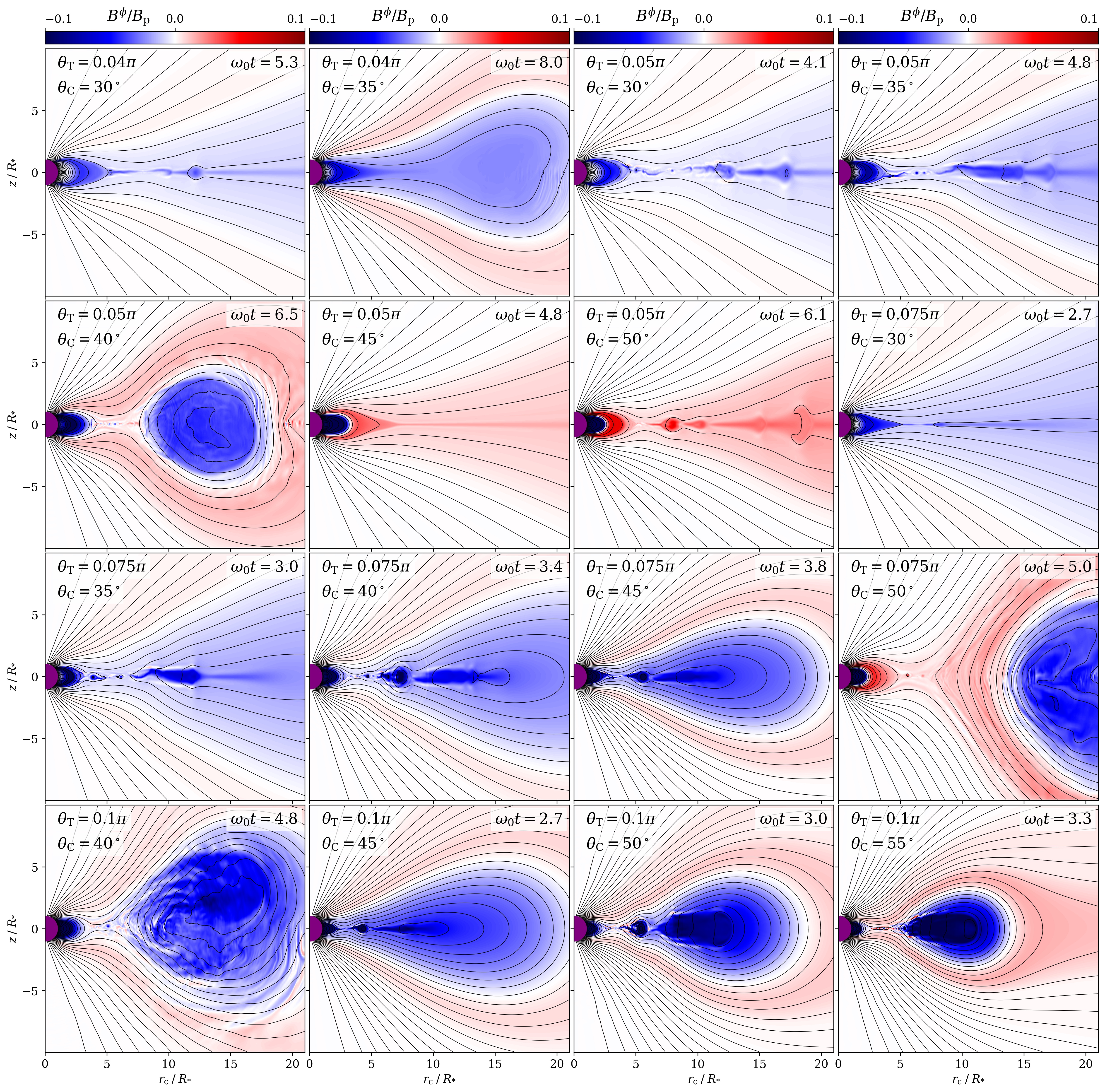}
 \vspace{-5pt}
 \caption{Snapshots of erupting magnetospheres for different profiles of the injected twist (varying the twisting disk colatitude $\theta_{\rm C}$ and extent $\theta_{\rm T}$) according to the setup described in Section~\ref{sec:axisymmetric}. All axisymmetric eruptions open up the dipole magnetosphere and form extended current sheets. Eruptions of flux bundles buried deep within the magnetosphere (bottom rows) are commonly accompanied by electromagnetic ejecta.}
\label{fig:2DModels}
\end{figure}

\subsection{Magnetospheric equilibrium conditions}
\label{sec:torusinstability}

A common starting point for dynamic models of the highly relativistic magnetospheres is given by equilibrium magnetic field configurations that balance magnetic pressure and magnetic tension. One important class of force-free equilibria, the minimum energy state, is the axisymmetric balance with vanishing toroidal current density.\footnote{Separable solutions to the spherical Grad-Shafranov equation, have a magnetic potential $\Psi(r,\theta)=g\left(r\right)h(\theta)$ with
\begin{align}
 \frac{r\sin\theta}{\sin\phi}\boldsymbol{\Delta}\left[\frac{\sin\phi}{r\sin\theta}\Psi\left(r,\theta\right)\right]=0.
 \label{eq:LaplacianGS}
\end{align}
Here, $\boldsymbol{\Delta}$ denotes the Laplacian and $\left(r,\theta,\phi\right)$ are spherical coordinates. Solutions to Equation~(\ref{eq:LaplacianGS}) are eigenfunctions of $\boldsymbol{\Delta}u=0$, where $u=\Psi\sin\phi/(r\sin\theta)$. They consist of Legendre polynomials of degree $l$.} In this Letter we examine the stability of dipole magnetospheres $\Psi_{l=1}\left(r,\theta\right)=B_0\sin^2\theta/r$ sheared by (not necessarily small) perturbations injected at the interior boundary. For slow injection rates, the magnetosphere will pass through a series of equilibria with nonzero toroidal current density \citep[recently derived as stationary solutions to a Grad-Rubin equation,][]{Stefanou2022}. Degenerate magnetospheres with dipolar conditions on the stellar surface but extended toroidal magnetic fields can become unstable and transition into a lower energy state \citep{Parfrey2013,Akgun2017,Akgun2018,Mahlmann2019}. The dynamic approach employed in axisymmetry by \citet{Parfrey2012,Parfrey2013} and for 3D setups in this work can capture the field line evolution during such instabilities consistently.

In the context of CMEs, several works examine the susceptibility of bent flux tubes to the torus instability -- a \emph{lateral} instability \citep{Kliem2006}. In magnetically dominated (but nonrelativistic) plasma, the balance of external Lorentz force and hoop force of a perturbed current ring can be used as a measure of stability against eruptions pushing through the external magnetic field $B_{\rm ex}$. \citet[][Chapter~4.7]{Bateman1978} defines the decay index $n=-R\times\text{d}\ln B_{\rm ex}/\text{d}R$ and gives the criterion $n\leq 3/2$ for the stability of a torus against toroidal expansion. For axisymmetric equilibrium configurations as discussed in Equation~(\ref{eq:LaplacianGS}) we find $n=2+l>3/2$ for a given multipole moment $l$. In other words, even though not studied in this work, magnetospheres with higher multipoles are likely to become more susceptible to the torus instability. The development of lateral instabilities shows some dependence on the field topology -- for example, the fractional length of the current ring within the domain \citep[e.g.,][]{Olmedo2010}. Still, the strongest and fastest CMEs typically have rapidly decaying external fields such that $n>3$ \citep[much larger than the range of uncertainty to the limits on the decay index $n$ discussed in the literature; see, e.g.,][]{Myers2017,Alt2021}.

\subsection{Dynamics of strongly twisted axisymmetric magnetospheres}

For comparison to the axisymmetric reference cases established by \citet{Parfrey2012,Parfrey2013} we first present a set of 2D simulations with $\left[N_r,N_\theta\right]=\left[64,200\right]$. In these models, we capture the magnetospheric twist induced by a cross section through the crustal perturbation described in Figure~\ref{fig:3DCompare} (panel c). Two adjoining rings in the northern hemisphere shear the footpoints of magnetic field lines differentially and thus mimic the dynamics that could be induced by fault lines running along the magnetar surface. The selected twist profiles are somewhat different from those found in the related literature \citep{Parfrey2012,Parfrey2013}, where field lines are sheared only in one direction on each side of the equator (see also discussion in Section~\ref{sec:discussion}). The double shearing profile, however, is comparable to the field line motion prescribed to the boundary in the context of solar CMEs \citep[][Figure 2]{Amari2003}. As we will show in the following, some CME-like field line dynamics can occur during eruptions also in the highly relativistic regime.

We reproduce the evolution stages of the differential shearing model for the example case of $\theta_{\rm C}=45^\circ$ and $\theta_{\rm T}=0.05\pi$ (Figure~\ref{fig:2DEvol}). In the initial phase, the magnetospheric twist grows linearly with time until it reaches a critical value (see also the calibration examples in Appendix~\ref{sec:eruptiondynamics}). During this stage, the global magnetospheric topology closely resembles that of a dipole, and the change of poloidal magnetic fields is small compared to the injected twist. The critical twist angle $\psi_{\rm crit}$ for which a twisted 2D magnetosphere opens up depends on the extent of the individual shearing region \citep[][Figure 14]{Parfrey2013}. In the presented setups, the outer layer of twisted magnetic field lines opens up first. The erupting flux bundle creates an extended current sheet that subsequently reconnects. Short-lived magnetic islands form and merge during the opening of the dipole geometry. While the twist increases further, the magnetosphere relaxes back to a dipolar, closed geometry. Finally, the lower layer of magnetic flux erupts, liberating much more energy than the first eruption. The strongly twisted states \citep[similar to degenerate Grad Shafranov equilibria, see][]{Akgun2018,Mahlmann2019} can eject blobs of coronal flux that propagate outward. 

In an auxiliary analysis of 2D magnetospheres, we vary the colatitude $\theta_{\rm C}$ and extent $\theta_{\rm T}$ of the twisting region to track the evolution of different magnetospheres during a time $\omega_0 t\approx 8$. In our analysis, we track the change in total magnetospheric energy and examine the magnetospheric dynamics when an event with notable energy dissipation occurs (i.e., a decrease in total energy or a change in the slope of the growth rate). Figure~\ref{fig:2DEruptions} summarizes the energetic imprint of 32 of such eruptions. The comparison between the total energy $e_{\rm twist}$ injected into the magnetosphere and the energy $e_{\rm diss}$ dissipated, for example, by violations of the force-free constraints shows an approximate scaling of $e_{\rm diss}/e_{\rm twist}=0.25$ (Figure~\ref{fig:2DEruptions}, right panel). Using this approximate scaling and assuming that Equation~\ref{eq:energybalance} can be used as a proxy for the amount of energy required to push the magnetosphere out of balance, one can estimate $e_{\rm diss}$ for different colatitudes $\theta_{\rm C}$ and fixed extension of the twist injection (Figure~\ref{fig:2DEruptions}, gray shaded region in the left panel). The estimate coincides well with the eruptions of the lower ring of the shearing region (indicated by blue colored bars). We note that the most energetic eruptions require sufficient magnetic energy to be stored deep inside the magnetospheres, along field lines close to the star. Therefore, the coincidence of the energy estimate (gray shaded region) with the spatial location of the lower ring (blue colored bars) is, indeed, a consistent finding. Equation~\ref{eq:energybalance} appears to be a valid rule-of-thumb estimate for the amount of energy that can be stored in the magnetosphere before an eruption occurs.

Figure~\ref{fig:2DModels} shows a collection of magnetospheric snapshots during eruption events for different twisting geometries (as indicated in each panel). As we discuss throughout this Letter, in the absence of other dissipative processes, axisymmetric models always drive the dipole topology to open up during periods of energy release. In contrast to the presented 3D magnetospheres that can dissipate energy locally in the strong guide field reconnection of helically unstable current tubes, the axisymmetry constraint only allows for two eruption scenarios. First, the twisted flux bundle can form a current sheet along the equator and drive reconnection events in a narrow region extending several tens of stellar radii (see also panel c of Figure~\ref{fig:2DEvol}). Second, a coronal flux ejection can disconnect from the twisted flux tube and bubble up. The current sheets forming during the disconnect of a magnetic bubble from a twisted flux bundle extend for a few stellar radii, are relatively short-lived, and produce significant plasmoid structures. We note that the bi-directional twist injected by the chosen twist geometry can be a reason for the appearance of coronal flux ejections. The opposite twist of the field lines covering a strongly sheared interior region can stabilize the magnetosphere against opening up on large scales, such as in the equatorial current sheet scenario.

In our extensive set of 2D axisymmetric models, all magnetospheres become unstable above $\psi_{\rm crit}$ \citep[][]{Parfrey2013}. However, we can identify two subscenarios of eruptions. First, the opening of the magnetosphere with an extended equatorial current sheet (see panel c, Figure~\ref{fig:2DEvol}). For the probed flux tubes enclosed by the colatitudes $\theta_{\rm E}\in\left[0.09\pi,0.23\pi\right]$ the critical angle $\psi_{\rm crit}$ for such events can be approximated (with respective standard errors) by 
\begin{align}
 \psi_{\rm crit}/(2\pi)=\left(0.26\pm 0.02\right)+\left(0.86\pm 0.04\right)\times \theta_{\rm E}.
\end{align}
Second, the expulsion of magnetic structures (as panel h, Figure~\ref{fig:2DEvol}). For larger colatitudes ($\theta_{\rm E}>0.17\pi$ in the studied models), critical twist values for this scenario are scattered with a mean and standard error of $\bar{\psi}_{\rm crit}/(2\pi)=0.8\pm 0.1$. The late-time evolution of such flux ejections is affected by dissipation via FFE constraint enforcement. Therefore, we do not follow the magnetospheric evolution further at this point. The field line topology displayed in Figure~\ref{fig:2DEvol} (panel h) is very similar to the erupting flux rope model for the solar photosphere discussed by \citet{Forbes1995}.

The collection of 2D eruptions presented in this paper shows a clear trend of $e_{\rm diss}/e_{\rm twist}\approx 0.25$ (right panel of Figure~\ref{fig:2DEruptions}). Here, $e_{\rm twist}$ is the total excess energy injected by the continuous twisting of the magnetosphere, and $e_{\rm diss}$ is the energy dissipated during an eruption by driving the magnetosphere to a force-free state. The measured fraction of dissipated energy further substantiates comparable estimates in the literature, such as $e_{\rm diss}/e_{\rm twist}\approx 0.1$ in \citet{Parfrey2013}. The uncertainty of $e_{\rm diss}/e_{\rm twist}$  estimates has to be reduced in future studies including relevant microphysics of dissipation. We theoretically estimate the energy liberated in axisymmetric eruptions by assuming a total available energy of $\mathcal{E}$ (Equation~\ref{eq:energybalance}). We find a good agreement between $e_{\rm diss}/e_{\rm twist}\times\mathcal{E}$, and the energy dissipated in simulations, as shown by the coincidence of blue lines and the gray shaded region in the left panel of Figure~\ref{fig:2DEruptions}.

\section{Numerical techniques}
\label{app:numericaltechniques}

\subsection{Perfect conductor boundaries}
\label{sec:perfectconductorboundary}

The surface of the central object is a boundary of utmost significance for the field line dynamics. For the application at hand, we find that implementing suitable boundary conditions on the level of Riemann problems is the most accurate \citep[see][]{Munz2000,Mahlmann2020b}. Studies of dynamics of magnetar magnetospheres commonly postulate a dipolar magnetic field anchored to the crust (similar to the one we display in Equation~\ref{eq:dipoleinit}). As the dipolar field does penetrate the boundary surface, we require the boundary conditions to fulfill 
\begin{align}
 \delta\mathbf{E}^*\times\mathbf{n}&=0,\\
 \delta\mathbf{B}^*\cdot\mathbf{n}&=0,
 \label{eq:boundaryapparentconductor}
\end{align}
where $\delta\mathbf{E}^*$ and $\delta\mathbf{B}^*$ are non-dipolar components of the electromagnetic fields in the boundary zone. Namely, we use $\delta\mathbf{B}^*=\mathbf{B}-\mathbf{B}_d$, as well as an analog for $\delta\mathbf{E}^*$. For derivations from the dipolar magnetosphere -- that do not need to be small -- the stellar surface thus appears as a perfect conductor, while the dipolar background is frozen into the surface layer for infinite times. We note that the derivation of suitable boundary routines for spherical surfaces emerges very naturally in the spherical version of our FFE code. Specifically, at the intercell face where the stellar surface is located, we set the following `left' (L) state of the Riemann problem (corresponding to the interior of the star):
\begin{align}
\Psi_L &= \Psi_R\\
\Phi_L &= -\Phi_R\\
\rho_L &= -\rho_R\label{eq:chargebound}\\
\mathbf{E}_L^* &= -\mathbf{E}_R^* + 2(\mathbf{E}_R^* \cdot \mathbf{\hat n}) \mathbf{\hat n}\\ 
\mathbf{B}_L^* &= +\mathbf{B}_R^* - 2(\mathbf{B}_R^* \cdot \mathbf{\hat n}) \mathbf{\hat n} ,
 \label{eq:BCs}
\end{align}
where $\mathbf{\hat n}$ is the unit radial vector (normal to the stellar surface) and R denotes the respective `right' state. For the highest accuracy, the reconstruction of $\delta\mathbf{E}_*$ and $\delta\mathbf{B}_*$ requires radial interpolations from the cell-centered fields in the magnetar crust to the boundary layer located at intercell faces. For this purpose, we recall the following proportionality relations for the dipole field: $B^\theta\propto r^{-4}$, $B^\phi \propto r^{-4}$, $E^r\propto r^{-2}$.

\subsection{3D domain and time step limitation}
\label{sec:filtering}

We extend the method introduced by \citet{Mahlmann2020b,Mahlmann2020c} to the specific challenges of 3D spherical meshes. The expressions for the proper grid spacing are
\begin{align}
\begin{split}
 \delta r &= \Delta r\\
 \delta \theta &= r \times \Delta \theta\\
 \delta \phi &= r\sin\theta \times \Delta \phi\\
\end{split}
\end{align}
The time step $\Delta t$ is then limited by the CFL factor $f_{\rm CFL}$,
\begin{align}
 \Delta t<f_{\rm CFL}\times\text{min}\left\{\delta r,\delta \theta,\delta \phi\right\}.
 \label{eq:CFLSpherical}
\end{align}
Operating close to the origin, the time step is limited by $\delta\theta$ if
\begin{align}
\begin{split}
 &\frac{\Delta r}{2} \times \Delta \theta = \delta\theta \lesssim \delta\phi=r\sin\theta \times \Delta \phi\\
\Longleftrightarrow\qquad n_\phi & \lesssim 4\times n_\theta \sin\theta \frac{r}{\Delta r}
\label{eq:limito}
\end{split}
\end{align}
Operating with an inner radial boundary that is sufficiently far away from the origin, as we commonly do for the simulation of surface-frozen dipole magnetospheres, the time step becomes limited by $\delta r$ if
\begin{align}
\begin{split}
 &\Delta r = \delta r \lesssim \delta\phi = r\sin\theta \times \Delta \phi\\
 \Longleftrightarrow\qquad n_\phi & \lesssim 2\pi\times \sin\theta \frac{r}{\Delta r}\equiv 2m
\label{eq:limite}
\end{split}
\end{align}
Independently from the actual number of grid points $n_\phi$, we effectively drive the validity of Equation~(\ref{eq:limite}) by damping all 1D Fourier modes that are of an order greater than $m$. Specifically, we filter such modes whenever $m < n_\phi/2$ by algebraically removing or damping all CFL-unstable modes in Fourier space. The simulated magnetospheres do not have significant deviations from axisymmetry at the coordinate singularity, and they are a well-suited application of such techniques. The Fourier filtering will be described in greater detail in a separate publication in the context of numerical simulations of the Einstein equations and general relativistic MHD (Ji et al 2023, in preparation). By applying such a radius and latitude-dependent filtering, along with the default choice of $\delta r\approx\delta\theta$ at the inner radial boundary, the condition prescribed by Equation~(\ref{eq:CFLSpherical}) relaxes to
\begin{align}
 \Delta t<f_{\rm CFL}\times\delta r.
\end{align}
The combination of filtering and angular rescaling (Equation~\ref{eq:angularscaling}) can increase $\Delta t$ by a factor $60-100$, making the simulations presented in this Letter computationally feasible.

As previously explored in \citet{Mignone2014} and \cite{Mahlmann2021}, the quality of the employed divergence cleaning techniques depends on the optimally chosen parameter
\begin{align}
 \alpha_{\sigma}=\frac{\kappa_\sigma}{c_\sigma}\Delta h\sim 1
 \label{eq:dampingparameter}.
\end{align}
Here, $\sigma$ represents the corresponding scalar potential ($\Psi$ for the constraint $\text{div}\mathbf{B}=0$; $\Phi$ for the constraint $\text{div}\mathbf{E}=\rho$), $\kappa_\sigma$ is the damping rate of numerical errors, $c_\sigma$ their advection rate, and $\Delta h=\text{min}\left[\Delta r,R_*\times\Delta\theta,R_*\sin\theta\times\Delta\phi \right]$ the locally determined resolution factor.

\begin{figure}
 \centering
 \includegraphics[width=0.48\textwidth]{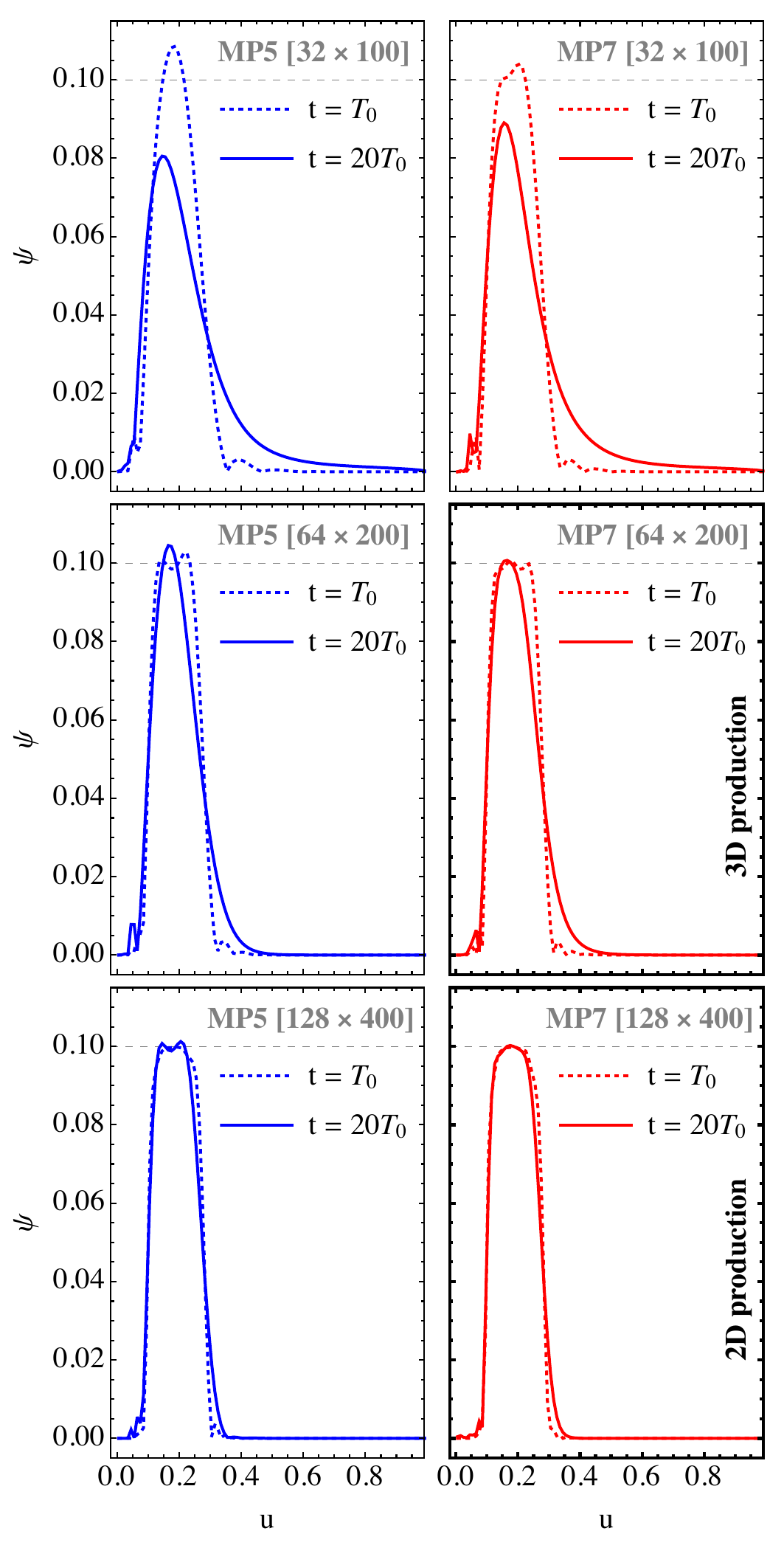}
 \vspace{-11pt}
 \caption{Profiles of the twist $\psi=|\phi_{\rm max}-\phi_{\rm min}|$ against the magnetic flux $u=\sin^2\theta$ for different resolutions and reconstruction schemes (as indicated).}
 \vspace{-11pt}
\label{fig:Diffusion}
\end{figure}

\section{Calibration benchmarks}
\label{app:calibationbenchmarks}

The numerical method employed throughout this work was extensively profiled by \citet{Mahlmann2020b,Mahlmann2020c}, with a focus on spherical geometries in \citet{Mahlmann2021}. In this section, we present the results of additional tests that focus specifically on the long-term stability of twisted field lines in the nonrotating dipole magnetosphere, with its delicate interior boundary (see Section~\ref{sec:perfectconductorboundary}).

\subsection{Diffusivity}
\label{sec:diffusivity}

\citet[][Section~4.2.2]{Parfrey2012} performed a convergence test that was specifically aimed at benchmarking the diffusivity for dipolar field geometries. Starting with a dipole configuration, selected field lines are displaced on the stellar surface, gradually increasing the toroidal magnetic field component by a slow twist until a time $t=T_0$. We reproduce the test cases of \citet[][Section~4.2.2]{Parfrey2012} by imposing an axisymmetric field line angular velocity
\begin{align}
 \omega\left(\theta\right)=\frac{\omega_{\rm C}}{1+\exp\left[-\kappa\left(g\left(\theta-\theta_{\rm C}\right)+\theta_{\rm T}\right)\right]}.
\end{align}
Here, $\theta_{\rm C}$ is the colatitude of the center of the twisting annulus, $\theta_{\rm T}$ is its angular half-width, and $\kappa=50$ denotes the decay rate at its edges. Furthermore, we use the factor $g=\text{sgn}\left(\theta_{\rm C}-\theta\right)$. The twisting is varied as a smooth profile in time $t$, returning to a vanishing angular velocity at the time $T_0$:
\begin{align}
 \omega_{\rm C}=\begin{cases} 
 \left(\omega_{\rm max}/2\right)\left[1-\cos\left(2\pi t/T_0\right)\right] & t\leq T \\
 0 & t>T_0 \\
 \end{cases}
\end{align}
We then monitor the subsequent stability of the induced twist over a total time interval of $20\times T_0$. As this test is quasi-axisymmetric, we conduct it on a 2D spherical grid, effectively employing three different resolutions $N_r=\left[32,64,128\right]$ and $N_\theta=\left[100,200,400\right]$. We also compare reconstruction schemes of different order \citep[fifth order, MP5; and seventh order, MP7;][]{Suresh1997} with the corresponding high-order finite difference approximation of source terms \citep[see][]{Mahlmann2020b}. 

Figure~\ref{fig:Diffusion} summarizes the quality of twist conservation and the effect of numerically induced diffusivity for an equilibrium configuration of the dipole magnetosphere. It can be directly compared to Figure~8 in \citet{Parfrey2012}. Increasing the grid resolution or the order of reconstruction reduces the diffusion of the established twist. The increased stability against inward leaking across strongly bent field lines is especially important for the simulations presented in this work. Already, the intermediate resolution of $\left[N_r,N_\theta\right]=\left[64,200\right]$ combined with MP7 reconstruction produces a very stable long-term conservation of the twist profile. An even higher resolution of $\left[N_r,N_\theta\right]=\left[128,400\right]$ suppresses the leakage even further. We conclude that the employed numerical method passes the convergence test prescribed by this dipole-specific twist diffusion benchmark.

\begin{figure}
 \centering
 \includegraphics[width=0.48\textwidth]{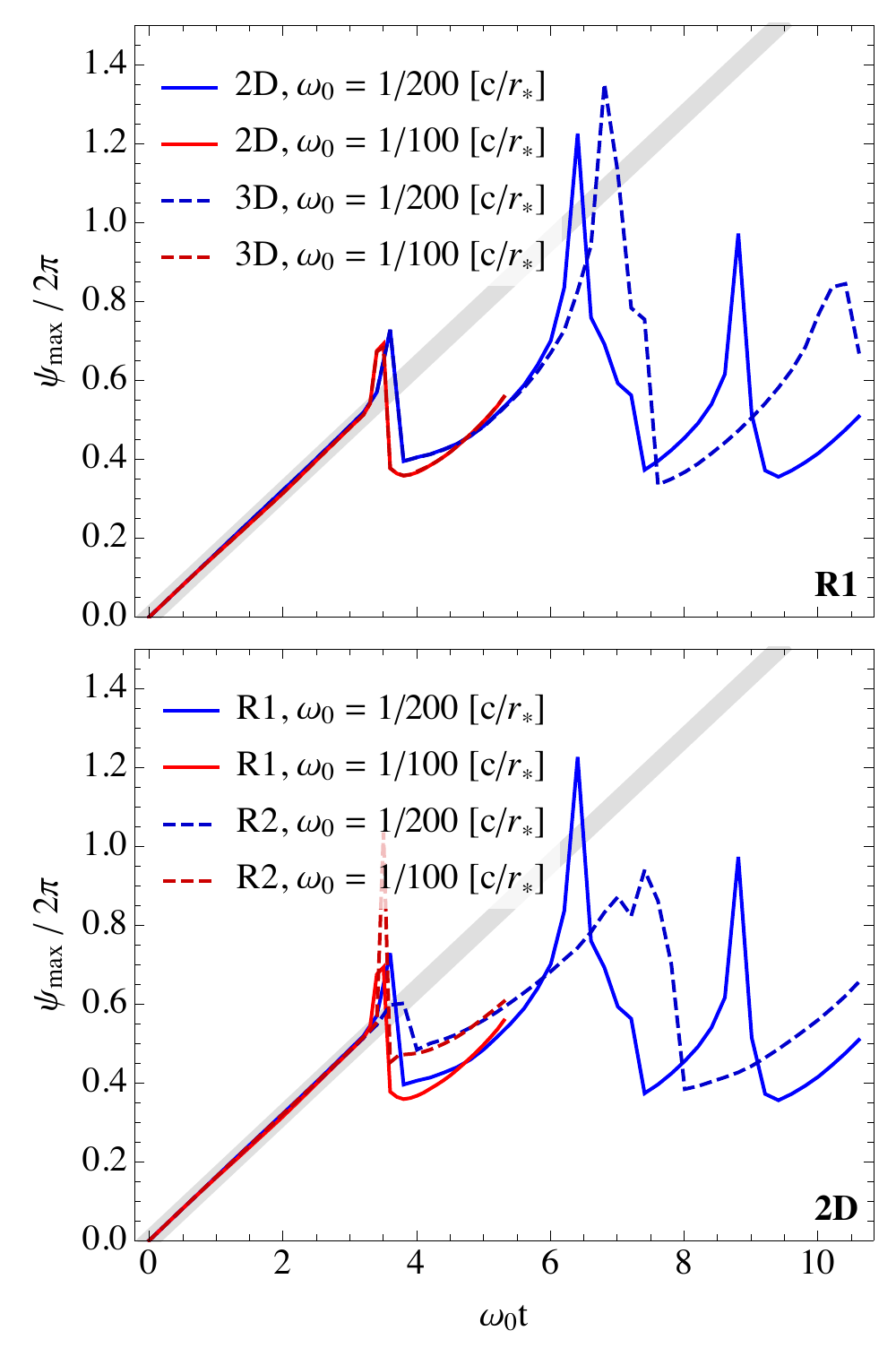}
 \vspace{-11pt}
 \caption{Evolution of the maximum twist during the continuous shear by a ring profile (\ref{eq:ringtwist}). The gray line denotes the expected linear growth of the twist. \textit{Top} panel: Comparison between 2D and 3D meshes for R1 for different base twist velocities $\omega_0$. \textit{Bottom} panel: Analog comparison, but between R1 and R2 for 2D meshes.}
 \vspace{-11pt}
\label{fig:Eruption}
\end{figure}

\subsection{Eruption dynamics}
\label{sec:eruptiondynamics}

In this section, we probe the 3D capacities of the employed spherical FFE infrastructure. While the previously examined diffusion test (Section~\ref{sec:diffusivity}) poses an inherently two-dimensional problem, the non-axisymmetric configurations of this work require full dimensionality. \citet{Parfrey2013} investigate the ring-like twist profile
\begin{align}
 \omega\left(\theta\right)=\frac{\omega_0}{1+\exp\left[\kappa\left(\left|\theta-\theta_{\rm C}\right|-\Delta\right)\right]}.
 \label{eq:ringtwist}
\end{align}
Here, $\theta_{\rm C}$ is the latitude of the ring's center, $\Delta$ its width, and we commonly choose $\kappa=50$. When sufficient twists are built up, it will trigger a reconnection event before transitioning into another twisting episode.

Building upon the previous section, we use MP7 reconstruction and two different meshes. First, we utilize a 2D configuration with two different resolutions, namely $N_r=\left[32,64\right]$ and $N_\theta=\left[100,200\right]$. Second, we probe the full dimensionality by conducting a fully 3D simulation with $N_r=32$, $N_\theta=100$, and $N_\phi=200$. We shall reference the lower of these two resolutions as R1, and the higher resolution as R2. The twist is induced and its dynamics are monitored during a time of approximately $1000$ light-crossing times of the central object. By comparing these simulations, we aim at exploring the following properties: a) The capturing of the initial twist induction phase in the 3D setup (with filtering, see Section~\ref{sec:filtering}). b) The first dynamic reconnection phase and its comparison between 2D/3D. c) The dependence of eruption dynamics on different resolutions.

Figure~\ref{fig:Eruption} assembles the time evolution of the key observable $\psi_{\rm max}$, namely, the maximum twist $\psi=|\phi_{\rm max}-\phi_{\rm min}|$ (see Section~\ref{sec:diffusivity}) per field line in the domain. The build-up phase of the twist ($t\leq 3.5\omega_0 t$) proceeds congruently between 2D and 3D meshes (top panel), as well as low (R1) and intermediate (R2) resolutions (bottom panel), following the theoretical expectation (as indicated by a thick gray line). The first breakup phase with significant twist rearrangement happens at the same time for all cases, with exceptional coincidence between 2D and 3D setups. While the late-phase ($t> 4\omega_0 t$) twist evolution can vary between different setups, the overall dynamics of shear and release persists, namely, a gradual build-up followed by a rapid decay. The differences in this phase are expected, as the reconnection phase inherently hosts nonideal regions (i.e., current sheets) and their evolution strongly depends on the numerical diffusion properties of the FFE method \citep{Mahlmann2021}. The very good coincidence between all models and the theoretical expectation in the build-up phase allows us to reliably calibrate the envisioned twist scenarios at low resolutions while relying on higher resolutions for the full development of field line instabilities.

\end{document}